\documentclass[amsmath,amssymb,prr,superscriptaddress,reprint,showpacs,longbibliography]{revtex4-1}
\usepackage{graphicx}
\usepackage{dcolumn}
\usepackage[colorlinks=true,linkcolor=blue,citecolor=blue,urlcolor=blue]{hyperref}
\usepackage[usenames,dvipsnames]{xcolor}
\usepackage{soul}
\usepackage{braket}
\usepackage{bm}
\usepackage{upgreek}
\usepackage{mathtools}
\usepackage{bbm}

\renewcommand{\vec}[1]{{\bf #1}}

\newcommand{\gaurav}[1]{\textcolor{red}{\textbf{[GC: #1]}}}
\newcommand{\aash}[1]{\textcolor{ForestGreen}{\textbf{[AAC: #1]}}}

\newtheorem{theorem}{Theorem}
\newtheorem{claim}{Claim}
\begin{document}

\title{A simple approach to characterizing band topology in bosonic pairing Hamiltonians }

% \title{Topological classification and characterization of quadratic bosonic systems}

%%%%%
%%%%%

\author{Gaurav Chaudhary}
\altaffiliation{Present address: Materials Science Division, Argonne National Laboratory, Lemont, IL 60439, USA}
\affiliation{James Franck Institute and Department of Physics, University of Chicago, Chicago, IL 60637, USA
}

\author{Michael Levin}
\affiliation{James Franck Institute and Department of Physics, University of Chicago, Chicago, IL 60637, USA
}

\author{Aashish A. Clerk}
\affiliation{Pritzker School of Molecular Engineering, University of Chicago, Chicago, IL 60637, USA}

\date{\today}


%%%%%%%%%%%%%%%%%%%%%%%%%%%%%%%%%%%%%%%%%%%%%%%%%%%%%%%%%%%%%%%%%%%%%%%%%%%%%%%%%%%%%%%%%%%%%%%%%%%%

\begin{abstract}
We revisit the problem of characterizing band topology in dynamically-stable quadratic bosonic Hamiltonians that do not conserve particle number.  We show this problem can be rigorously addressed by a smooth and local adiabatic mapping procedure to a particle number conserving Hamiltonian.  In contrast to a generic fermionic pairing Hamiltonian, such a mapping can always be constructed for bosons.  Our approach shows that particle non-conserving bosonic Hamiltonians can be classified using known approaches for fermionic models.  It also provides a simple means for identifying and calculating appropriate topological invariants.  We also explicitly study dynamically stable but non-positive definite Hamiltonians (as arise frequently in driven photonic systems).  We show that in this case, each band gap is characterized by two distinct invariants.  
% We study the gapped phases of dynamically stable quadratic bosonic Hamiltonians and show that an arbitrary gapped phase can be adiabatically mapped to a particle number conserving Hamiltonian. Our mapping shows that any gapped topological phase of a dynamically stable quadratic bosonic system inherits its topology from Altland-Zirnbauer ten-fold classification.
% By explicitly mapping the bosonic system to number conserving Hamiltonian, we provide a convenient method to calculate the topological invariants using the well known expressions for the equivalent fermionic problem.
% We further prove that the zero energy gap for the positive definite system is always topologically trivial.
% Interestingly, topologically non-trivial phase in the zero energy gap can be induced by relaxing positive definite condition. 
% Our mapping procedure is more intuitive than existing literature and reveals interesting features in the topological invariants of the bosonic system compared to their fermionic counterparts.
\end{abstract}
%%%%%%%%%%%%%%%%%%%%%%%%%%%%%%%%%%%%%%%%%%%%%%%%%%%%%%%%%%%%%%%%%%%%%%%%%%t%%%%%%%%%%%%%%%%%%%%%%%%%%

\maketitle

%%%%%%%%%%%%%%%%%%%%%%%%%%%%%%%%%%%%%%%%%%%%%%%%%%%%%%%%%%%%%%%%%%%%%%%%%%%%%%%%%%%%%%%%%%%%%%%%%%%%

\section{Introduction\label{Sec:Intro}}
%\ml{From my point of view, there are 3 levels of classification of quadratic systems: (1) classification of symmetries, i.e. AZ scheme; (2) classification of TI's in a given symmetry class, i.e. the ``periodic table of topological insulators and superconductors''; (3) band invariants and methods to analyze concrete systems in a given symmetry class. Our work is mostly relevant to (3), and maybe also (2) to a lesser extent. I'm worried that we may give the impression that our work is closer to (1) and (2) with all the references to classification and AZ, etc. Let's discuss.}

Topological phases of matter lie at the heart of modern condensed matter physics~\cite{Thouless1982,Haldane1988,Kane2005,Bernevig2006}.
% The widespread interest in their observation, manipulation, and classification stems from
%They are interesting both for their novel underlying fundamental physics~\cite{Witten2016}, as well as for possible applications to quantum information processing~\cite{Kitaev2003,Nayak2008,Alicea2011,Barends2014}.
In solids, topology is commonly encountered via  electronic properties, such as quantized Hall conductivity~\cite{Klitzing1980,Tsui1982}, electronic bands of topological insulators~\cite{Kane2005,Bernevig2006,Hsieh2008}, and Bogoliubov quasiparticle bands of topological superconductors~\cite{Read2000,Kitaev2001,Mourik2012}.
These are examples of topological phases of free fermions. In the absence of % constraints imposed by 
additional crystalline symmetries, 
%and strong interactions
the topological classification of gapped phases of quadratic fermionic Hamiltonians is described by the periodic table of topological insulators and superconductors~\cite{schnyder2008classification, kitaev2009periodic, schnyder2009classification, ryu2010topological}, which is closely related to the Altland-Zirnbauer (AZ) classification of random matrices~\cite{Altland1997}. The guiding principles behind this classification scheme are the presence or absence of  time reversal, charge conjugation, and chiral symmetries along with the dimensionality of the system.

Recently, there has been a surge of interest in studying topological phenomena in bosonic systems. 
In solids, topological bandstructures of  phonons~\cite{Prodan2009,Kane2013,Peano2015,Yang2015}, magnons~\cite{Kim2016,Nakata2017,Laurell2018,Lu2018,Lee2018,Kondo2019,Kondo2019a}, and excitons~\cite{Wu2017,Hu2018} have been studied. Bosonic topology can also be engineered in cold atoms~\cite{Atala2013,Jotzu2014,Goldman2014,Aidelsburger2014,Stuhl2015}, photonic systems~\cite{Haldane2008,Khanikaev2012,Hafezi2013,Rechtsman2013,Lu2014,
Karzig2015,OzawaRMP2019} and mechanical metamaterials~\cite{Kane2013,Chen2014,Nash2015,Khanikaev2015,Susstrunk2016,Prodan2017}.
%, which provide novel route to control and manipulate these phases.  
Given this interest, a simple approach to bosonic band topology would be extremely valuable.  Stated explicitly, we would like to understand when the bandgaps of two different quadratic bosonic Hamiltonians are topologically equivalent (i.e.~can one smoothly interpolate from one to the other without closing the gap of interest?).  
%In this work, we focus on bosonic band topology, which is often the quantity of interest for photonic and phononic systems.

%\aash{Should perhaps already make it clear that we are interested primarily in classifying band topology, and not the topology of some many-body state.  Stress that for non-conserved particles like phonons or photons, this is the thing that is experimentally relevant.}
%\gaurav{I have added a sentence here. Although, The second half of first paragraph does indicate the same.}

For quadratic, particle-number conserving bosonic Hamiltonians, the band topology problem is largely understood, since the single-particle wave equation is identical for fermions and bosons.  The bosonic band topology problem is thus identical to the well-studied fermionic problem in this case. However, in the presence of terms that break particle number conservation, the situation changes drastically~\cite{Shindou2013,Engelhardt2015,Bardyn2016,Peano2016a}.
% Topologically non-trivial phases can be induced in an otherwise a trivial bosonic system through introduction of  pairing terms
In these particle non-conserving cases, the band structure of the bosonic system is obtained by diagonalizing its non-Hermitian dynamical matrix.
% Non-Hermiticty is ubiquitous in nature and leads to exotic phenomena  such as unconventional transmission and reflection~\cite{Feng2012}, parity-time symmetry~\cite{Bender1998,Guo2009,Rueter2010,Lin2011}, and non-Hermitian skin effect~\cite{Yao2018,Yao2018a}.
% Generally, non-Hermitian matrices have complex eigenspectrum, which leads to dynamical instabilities.
This makes the bosonic band topology problem at first glance possibly richer and more complex than the standard fermionic problem.  One approach is to view the bosonic problem as a particular example of the far more general problem of classifying topology of an arbitrary  non-Hermitian Hamiltonian~\cite{Gong2018,Kawabata2019,Lee2019,Zhou2019}. 
A symmetry-based approach to this problem (involving 38 distinct symmetry classes) was recently presented in Refs.~\cite{Kawabata2019,Zhou2019}.

% \michael{Add comment motivating approach: we can leverage fermionic results for bosonic case, avoid complicated classification schemes.}

\begin{figure}[tb]
%  \begin{tabular}{c c}
%    (a) \includegraphics[width=0.45\textwidth]{figures/E_DOS_1.png} & (b) \includegraphics[width=0.45\textwidth]{figures/E_DOS_2.png} \\
%    (c) \includegraphics[width=0.45\textwidth]{figures/E_DOS_3.png} & (d) \includegraphics[width=0.45\textwidth]{figures/E_DOS_4.png}
%  \end{tabular}
  \includegraphics[width=0.3\textwidth]{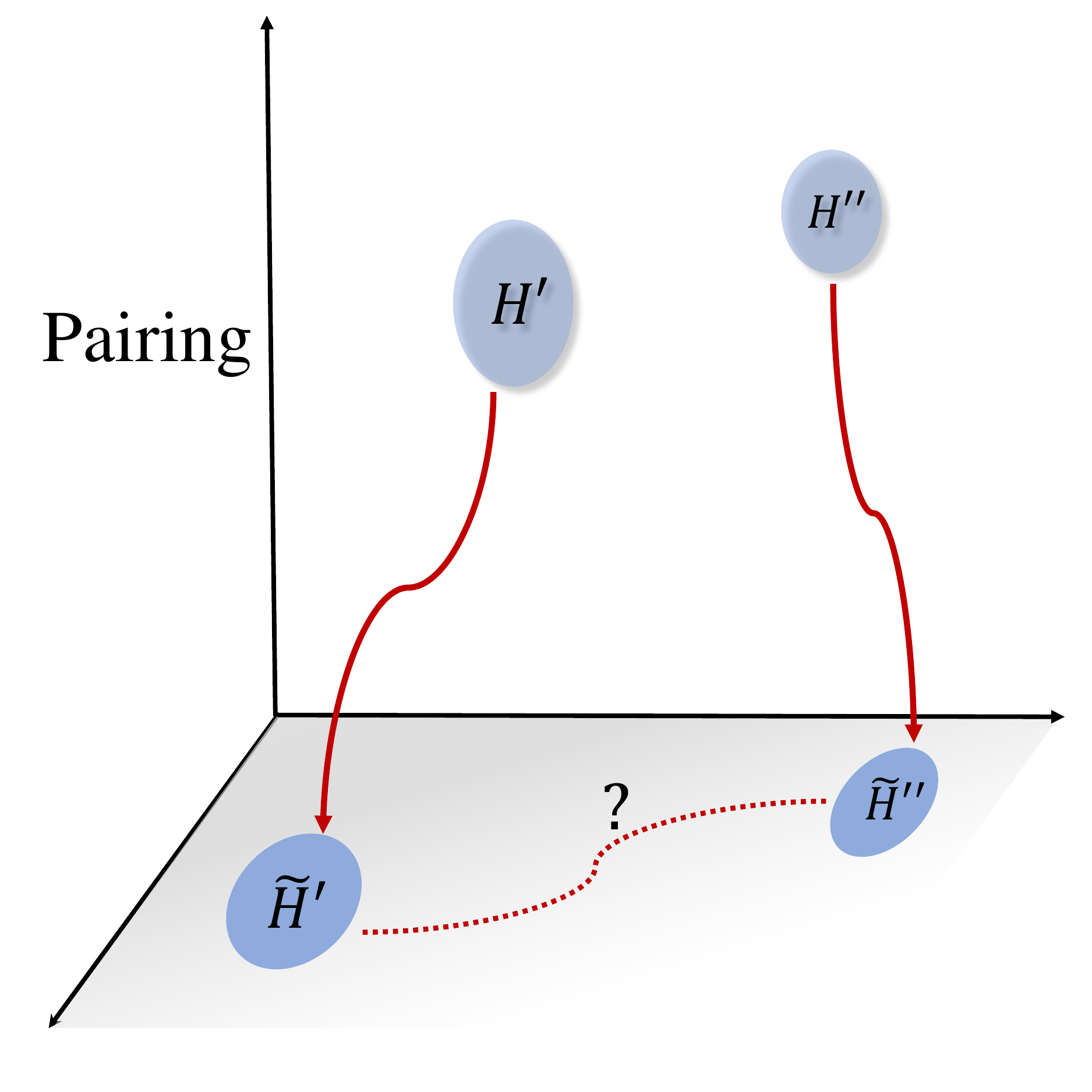}
  \caption{\label{Fig:Schematic}
  Schematic representation of the topology preserving map between dynamically stable bosonic Hamiltonians and number conserving Hamiltonians. The shaded horizontal plane represents the space of all number conserving Hamiltonians. Our map, denoted by the solid curved lines, connects a general dynamically stable Hamiltonian $H$ to a number conserving Hamiltonian $\tilde{H}$. Two Hamiltonians, $H'$ and $H''$, can be adiabatically connected if and only if the corresponding number conserving Hamiltonians $\tilde{H}'$ and $\tilde{H}''$ can be adiabatically connected. }
\end{figure}

Here, we describe a simpler way to analyze the topology of quadratic bosonic Hamiltonians with pairing terms: instead of making recourse to new classification schemes, one can directly leverage the well-known results for particle conserving fermionic Hamiltonians. We present a simple, topology-preserving mapping between dynamically-stable bosonic pairing Hamiltonians and particle conserving Hamiltonians
(see Fig.~\ref{Fig:Schematic}).
This mapping provides a direct and unambiguous method for determining whether band gaps in two different bosonic systems are topologically equivalent, and allows one to connect to the standard periodic table of topological insulators and superconductors~\cite{schnyder2008classification, kitaev2009periodic, schnyder2009classification, ryu2010topological}. 
% This mapping also provides several important physical insights, and serves as a useful computational tool for calculating invariants.
In addition to providing a clear physical picture, our approach helps address  several important open questions.  
%about topology in bosonic pairing Hamiltonians.  
In particular:
\begin{itemize}
    \item Can the band topology of a bosonic pairing Hamiltonian  be distinct from what is possible with particle conserving Hamiltonians?  
    \item Is there a simple and explicit recipe for calculating topological invariants for these bosonic pairing systems?  How do these relate to standard invariants in particle-conserving systems?  
    % \item SOMETHING ELSE?
    % \item Can the gap at zero energy in a stable bosonic system ever be topologically non-trivial?  Previous work demonstrated that for positive definite Hamiltonians, this is not possible \cite{Shindou2013}.  However, the situation for more exotic Hamiltonians that are stable but not positive definite is unclear.  Such systems can be realized in driven photonic or phononic setups~\cite{Peano2016}.
\end{itemize}

 The mapping and analysis we present has additional virtues.
 Our mapping explicitly preserves any symmetry of the original pairing Hamiltonian that commutes with total particle number.  As a result, more refined topological classifications sensitive to additional symmetries (e.g.~crystalline symmetries) can be addressed using our equivalent particle-conserving Hamiltonian.  Further, we show that the topology of a particular band gap can be fully calculated by only considering the particle-like parts of the relevant Bogoliubov quasiparticle wavefunctions.  We also discuss why this interpolation procedure from particle non-conserving to conserving models cannot be accomplished generically for fermionic pairing Hamiltonians.

Our work extends and complements previous studies of bosonic topology based on mapping of positive-definite models to fermionic models \cite{Susstrunk2016,Lu2018,Lein2019,Xu2020}.
An issue with some of these mappings is that they simply generate a fermionic model which is isospectral to the original bosonic system; as discussed further in Sec.~\ref{subsec:previous}, this is not always sufficient to address questions of classification. 
%In contrast, our approach provides a more direct means for classifying bosonic systems using the standard AZ classes. 
Our work is also more general, in that it also applies to bosonic pairing Hamiltonians that are not positive definite but are nonetheless dynamically stable.  While such Hamiltonians do not arise in typical condensed matter situations (e.g.~spin-wave Hamiltonians), they are common in driven photonic systems, where one is working in a rotating frame determined by an external pump frequency.  We show that for these non-positive definite cases, it is physically meaningful to assign two independent invariants to each band gap (that characterize the particle and hole like edge states expected in a system with open boundary conditions).
Further, there is no guarantee that there exists a topologically trivial bandgap (e.g.~the gap at zero energy).  Both these features are in stark contrast to the positive definite case. 
%In that case, there is at most a single non-zero invariant for each bandgap (i.e.~no co-existence of particle-like and hole-like edge states), and the bandgap at zero energy is always guaranteed to be trivial \cite{Shindou2013}.  
The existence of two non-zero invariants in a single gap can have a striking physical consequence:  in such cases, the system can become dynamically unstable when a boundary is introduced (a phenomenon that has been noted in several specific models previously \cite{Barnett2013,Barnett2015,Peano2016a}).  
Note that our approach strongly differs from previous studies of general non-positive definite bosonic pairing Hamiltonians.  
Ref.~\cite{Peano2018a} also addresses topology of such systems, but did not connect to particle conserving models. 
Ref.~\cite{Flynn2020} does introduce such a mapping; however, unlike our approach, this map is not guaranteed to be local, nor can it be used in general to address topology.

The article is organized as follows. In Sec.~\ref{Sec:background}, we review some basic properties of quadratic bosonic systems. 
In Sec.~\ref{Sec:result} we present our mapping between pairing and number-conserving Hamiltonians, as well the main implications that follow.  In Sec.~\ref{Sec:interp}, we prove our main results by constructing an explicit interpolation between a general quadratic bosonic system and its number conserving partner. 
In Sec.~\ref{Sec:implication}, we discuss some important implications of our main result for the classification and analysis of quadratic bosonic Hamiltonians.
In Sec.~\ref{Sec:Proc}, we discuss some additional details of our mapping procedure.
In Sec.~\ref{Sec:num}, we present some numerical results.
We conclude in Sec.~\ref{Sec:dis}. 
Some technical details are presented in the appendices.

\section{Background: Bandstructure of a general quadratic bosonic Hamiltonian\label{Sec:background}}

Consider a general $d$-dimensional lattice bosonic system with $N$ sites per unit cell, described by a quadratic Hamiltonian with short-ranged hopping and pairing terms.  Assuming periodic boundary conditions and translational invariance, the momentum space Hamiltonian has the general form
\begin{align}\label{Eq:system_ham}
      H_B &= \sum_{\vec{k}} \sum_{n,n'=1}^N t_{n,n'}(\vec{k})\, b^{\dagger}_{n,\vec{k}} b_{n',\vec{k}}\notag\\
     &+\, \frac{1}{2}\left ( \sum_{\vec{k}} \sum_{n,n'=1}^N\lambda_{n,n'}(\vec{k})\, b^{\dagger}_{n,\vec{k}}b^{\dagger}_{n',-\vec{k}}\,+\,h.c. \right) \, ,
\end{align}
where $b_{n,\vec{k}}$ annihilates a boson on sublattice $n$ and momentum $\vec{k}$.
% The Hamiltonian can be brought to the Bogoliubov de-Gennes (BdG) form, which presents a convenient method to diagonalize the system. 
Introducing the Nambu vector  $\phi_{\vec{k}} = (b_{1,\vec{k}},\, ...,\,b_{N,\vec{k}},b^{\dagger}_{1,-\vec{k}}\, ...,\,b^{\dagger}_{N,-\vec{k}}  )^T$, we can write $H_B$ compactly in the Bogoliubov de-Gennes (BdG) form
\begin{subequations}
\begin{align}
    & H_B = \frac{1}{2}\sum_{\vec{k}} \phi^{\dagger}_{\vec{k}}\, h_B(\vec{k})\, \phi_{\vec{k}}\, , \\
    & h_B(\vec{k}) = \begin{pmatrix}
    t(\vec{k}) & \lambda(\vec{k})\\
    \lambda^{\dagger}(\vec{k}) & t^{\ast}(-\vec{k})\\
    \end{pmatrix}\, .
\end{align}
\end{subequations}
where $t(\vec{k})$ and $\lambda(\vec{k})$ are $N\times N$ matrices related to hopping and pairing respectively.
% Notice that compared to the fermionic BdG matrix, the lower diagonal block representing 
% the ``hole'' kinetic energy comes with the same `\textit{overall sign}' as the upper diagonal particle kinetic energy block. 
% As a result, the bosonic BdG matrix can be positive definite, unlike the fermionic BdG matrix, which is always particle-hole symmetric and never positive definite.

We are interested in the band structure of our system. As in fermionic problems, this band structure can be defined by diagonalizing $H_B$ via a Bogoliubov transformation. That is, we rewrite $H_B$ as
\begin{align}
     & H_B = \sum_{\vec{k}}\sum_{m=1}^N \omega_{m}(\vec{k})\, A^{\dagger}_{m,\vec{k}} A_{m,\vec{k}}.
     \label{eq:HDiagonal}
\end{align} 
where $A_{m,\vec{k}}$ are canonical, independent bosonic annihilation operators of the form
\begin{align}
   & A_{m, \vec{k}} = \sum_{n=1}^N (u_{mn}(\vec{k})\, b_{n,\vec{k}} + v_{mn}(\vec{k})\, b_{n, -\vec{k}}^\dagger)\, ,
\label{Eq:mode_annihilation}    
\end{align} 
and where $u_{mn}$ and $v_{mn}$ are complex coefficients. The energies $\omega_{m}(\vec{k})$ define the band dispersions of our system (or more precisely the band dispersions of the $N$ bands with positive symplectic norm, as we explain below).

It is well known that some bosonic pairing Hamiltonians cannot be diagonalized as in Eq.~(\ref{eq:HDiagonal}). In this paper we do not consider such systems. Instead, we restrict ourselves to ``dynamically stable'' Hamiltonians $H_B$ which are characterized by two properties: (i) $H_B$ is diagonalizable and (ii) the energies $\omega_m(\vec{k})$ are all real. Note that this notion of dynamic stability does \emph{not} require $H_B$ to be positive definite. Throughout this paper we assume our system is dynamically stable (and hence diagonalizable), but not necessarily positive definite.  

%where the time evolution generated by $H_B$ gives rise to exponentially growing modes~\cite{Moiseyev2011}.

A direct route for obtaining the above band structure is to consider the Heisenberg equations of motion generated by $H_B$, namely
$\dot{\phi} = -i\tau_z h_B\, \phi$. The matrix on the right hand side defines the ``dynamical matrix'' 
\begin{align}
D (\vec{k}) \equiv \tau_z h_B(\vec{k}). 
\end{align}
To obtain the band structure, one needs to find the eigenvalues and (right) eigenvectors of the dynamical matrix $D (\vec{k})$:
\begin{align}
    & D (\vec{k})\,|\vec{k}_{j}\rangle = \omega_{j}(\vec{k})\, |\vec{k}_{j}\rangle.
\label{Eq:dynamical}
\end{align}
Here $j$ indexes the $2N$ eigenvalues/vectors of $D(\vec{k})$, and $|\vec{k}_{j}\rangle$ denotes a $2N$ component vector. Our dynamical stability assumption is equivalent to assuming that the dynamical matrix $D(\vec{k})$ is diagonalizable and has real eigenvalues. Given this assumption, it is not hard to show that the eigenvectors $|\vec{k}_{j}\rangle$ can always be chosen so that (i) they are orthogonal with respect to the symplectic inner product $\langle \vec{k}_{i} |\, \tau_z \,|\vec{k}_{j}\rangle$, and (ii) they have strictly positive or strictly negative symplectic norm $\langle \vec{k}_{j} |\, \tau_z \,|\vec{k}_{j}\rangle$.

%The matrix $D$ is non-Hermitian and hence in the general case, the eigenvalues $\omega_j$ are complex.  However, in the dynamically stable case we focus on, all eigenvalues are (by definition) real.
%We also assume throughout that $D$ is diagonalizable, hence exclude cases of marginal stability~\cite{Peano2016}.
%The eigenvectors $|\vec{k}_{j}\rangle$ in this case either have a positive or negative symplectic norm $\langle \vec{k}_{j} |\, \tau_z \,|\vec{k}_{j}\rangle$.

%We label the $N$ positive norm (negative norm) states by $j = (m,+)$ ($j = (m,-)$).  
With these properties in mind, we can now explain the connection between Eq.~(\ref{Eq:dynamical}) and Eq.~(\ref{eq:HDiagonal}): the energies $\omega_m(\vec{k})$ in Eq.~(\ref{eq:HDiagonal}) are simply the $N$ eigenvalues of $D(\vec{k})$ that correspond to \emph{positive}-norm eigenvectors $|\vec{k}_{j}\rangle$. Likewise, the corresponding Bogoliubov coefficients $u_{mn}$ ($v_{mn}$) are given by the first-$N$ (last $N$) components of the vector $\langle \vec{k}_m |\tau_z$. (Here, $|\vec{k}_m\rangle$ has positive symplectic norm normalized to 1).
From now on we refer to the eigenvectors of $D(\vec{k})$ with positive (negative) norm as ``particle-like'' (``hole-like'') modes.

% symplectic norm, $\langle \vec{k}_{m,\sigma} |\, \hat{\tau}_z \,|\vec{k}_{m,\sigma}\rangle \neq 0$, which can be positive or negative.  As is standard, we normalize these eigenvectors to have norm $\pm 1$  For our fully stable case, there are $N$ eigenvectors with norm $+1$ (which we label by $j \rightarrow (m,+))$, and $N$ eigenvectors with norm $-1$ (which we label by $j \rightarrow (m,-))$.
% dynamical-mode eigenvectors corresponding to distinct eigenvalues are orthogonal with respect to the symplectic product $\langle \vec{k}_{m,\sigma} |\, \hat{\tau}_z \,| \vec{k}_{m',\sigma'} \rangle = 0$.  Further, eigenvectors corresponding to real energies have a nonzero 

Another general fact that will be useful in what follows is that the dynamical matrix has an effective ``particle-hole'' symmetry which guarantees that modes come in pairs with opposite wave vectors $(\vec{k}, -\vec{k})$ and opposite frequencies $(\omega, -\omega)$. This symmetry follows from the identity
\begin{align}
    (\kappa \tau_x) D(\vec{k}) (\kappa \tau_x) = -D(-\vec{k})
\end{align}
%\begin{align}
%   & D(-\vec{k}) \, (\kappa \tau_x |\vec{k}_j\rangle ) = %-\omega_j(\vec{k})\, (\kappa \tau_x |\vec{k}_j\rangle)\, ,
%\label{Eq:phsymmetry}   
%\end{align} 
where $\kappa$ represents complex conjugation. Each pair of eigenvectors is of the form $(|w\rangle , \kappa \tau^x |w \rangle)$, and always consists of one ``particle-like'' state and one ``hole-like'' state.    

Given this redundancy in the spectrum of the dynamical matrix, there are two ways to think about the bandstructure of our bosonic problem:
\begin{enumerate}
    \item One can just focus on the energies of the particle like states, as one can write the diagonalized second quantized Hamiltonian entirely in terms of these states, c.f.~Eq.~(\ref{eq:HDiagonal}).  We thus have $N$ particle-like bands.  
    \item  One can instead consider all eigenvalues of the dynamical matrix.  We thus have $2N$ bands (half of which are particle, the other half holes), and this band structure has an effective particle-hole symmetry (i.e.~if $\omega[k]$ is a band energy, then so is $-\omega[-k]$).  This kind of ``doubled'' bandstructure
    is sketched in Fig.~\ref{Fig:Flattening}(a).
\end{enumerate}
While both these viewpoints provide equivalent descriptions, we will use the latter notion, as it is especially helpful in the case of non-positive definite (but dynamically stable) Hamiltonians.  In this general case, there is no simple way to immediately distinguish particle and hole bands by the sign of the energy:  both can exist at either positive or negative energies.  

\section{Main result\label{Sec:result}}

%\aash{Weren't we going to say something about symmetry early in Sec. III?  i.e. start by saying that we start by ignoring additional symmetries, then show later that these can be included in our procedure?}

%\aash{I think we should say something briefly in words about band-flattening (i.e. start with the general Hamiltonian, focus on a particular gap, flatten spectrum in a way to not impact that gap.  If we don't say something, Eq. 7 might seem like an odd starting point.}
%\gaurav{Check the changes}

Our main result is a topology preserving mapping defined on quadratic bosonic Hamiltonians. This mapping associates to every dynamically stable bosonic Hamiltonian a corresponding particle number conserving bosonic Hamiltonian that has the same topology with respect to a given spectral gap. 

We first explain the mapping in the simplest case: bosonic systems without any symmetry constraints apart from lattice translation symmetry. Later we will discuss how additional symmetries, such as time reversal or crystalline symmetries, can be included in our approach.

\begin{figure}[tb]
%  \begin{tabular}{c c}
%    (a) \includegraphics[width=0.45\textwidth]{figures/E_DOS_1.png} & (b) \includegraphics[width=0.45\textwidth]{figures/E_DOS_2.png} \\
%    (c) \includegraphics[width=0.45\textwidth]{figures/E_DOS_3.png} & (d) \includegraphics[width=0.45\textwidth]{figures/E_DOS_4.png}
%  \end{tabular}
  \includegraphics[width=0.45\textwidth]{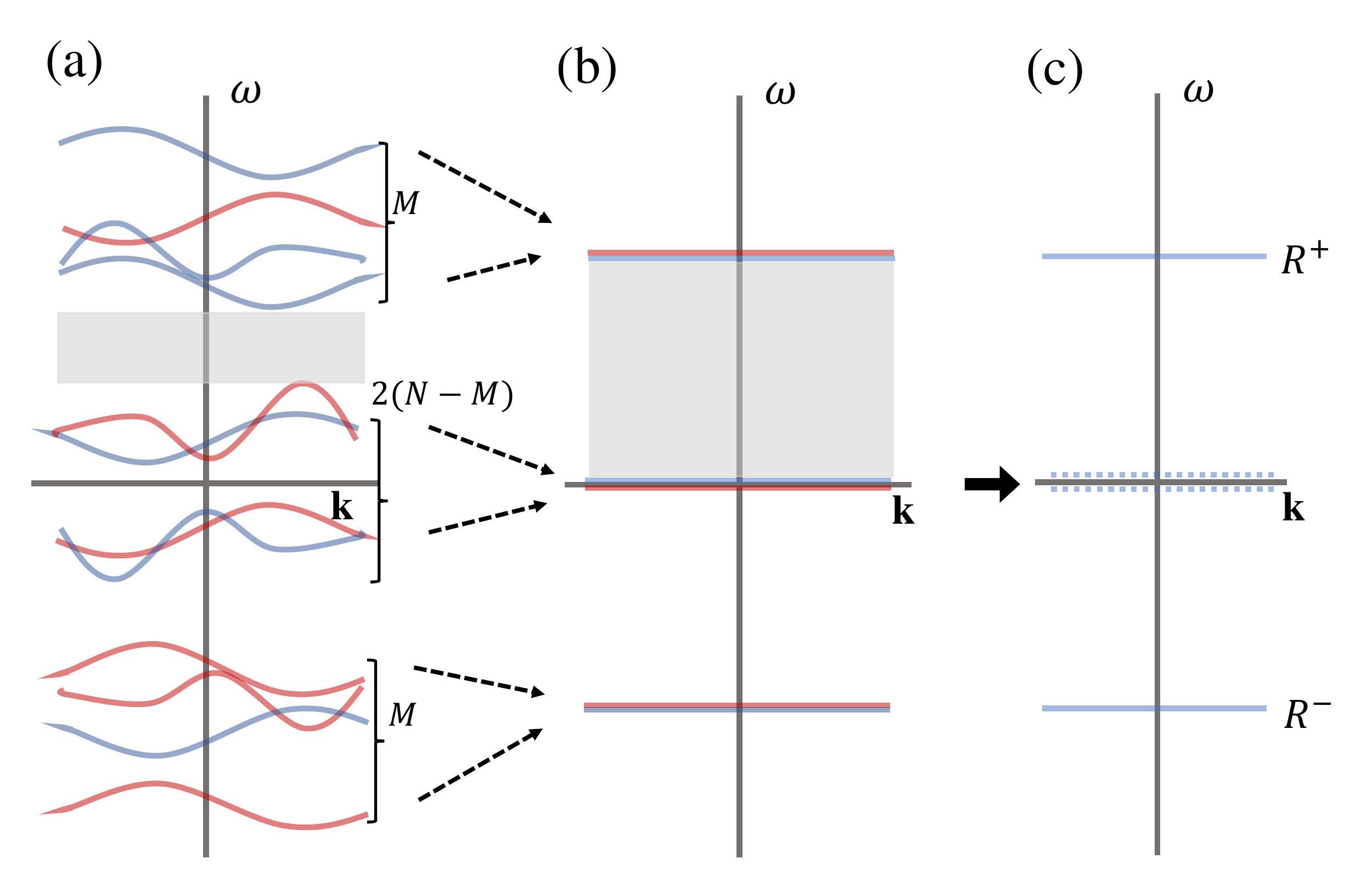}
  \caption{\label{Fig:Flattening}
  Band flattening: (a) A qualitative representation of the spectrum of dynamical matrix $D$ of original system, (b) flattening of the dynamical matrix, (c) the flattened spectrum of second quantized Hamiltonian $H$.  Here the blue and red bands represent the particle and hole-like states respectively and the energy gap of interest is presented by shaded gray region. In (c) the dashed bands at zero energy are not explicitly present in the second quantized Hamiltonian. }
\end{figure}

To begin, consider a general dynamically stable bosonic Hamiltonian of the form given in Eq.~(\ref{Eq:system_ham}). As shown in Fig.~\ref{Fig:Flattening} (a), we assume that the corresponding dynamical matrix has a spectral gap around some frequency $\pm \omega_0$, with $2M$ bands with frequencies $|\omega|> \omega_0$ and $2(N-M)$ bands with frequencies $|\omega| < \omega_0$. Here, $M$ is an integer with $0 \leq M \leq N$.
%\aash{Should we refer to Fig. 1 here, and also label $M$ and $N-M$ in the figure?}

In order to apply our mapping, we first perform a preprocessing step where we flatten the spectrum of the Hamiltonian of interest around $\pm \omega_0$ as shown in Fig.~\ref{Fig:Flattening} (b). In this flattening procedure, the $M$ bands of the dynamical matrix with frequencies $\omega > \omega_0$ are mapped to flat bands with frequency $\omega = 1$. Likewise, the $M$ bands with frequencies $\omega < - \omega_0$ are mapped to flat bands with frequency $\omega = -1$. Finally, the $2(N-M)$ bands with frequencies $|\omega| < \omega_0$ are mapped to flat bands with frequency $\omega = 0$. This spectral flattening step is schematically shown in Fig.~\ref{Fig:Flattening} and discussed in more detail in Sec.~\ref{Sec:Proc}. 

After flattening, we then write the resulting second quantized Hamiltonian in the diagonalized form
\begin{align}
    & H = \sum_{\vec{k}} \left(\sum_{m=1}^{R_+} A_{m,\vec{k},+}^\dagger A_{m,\vec{k},+} - \sum_{m=1}^{R_-} A_{m,\vec{k},-}^\dagger A_{m,\vec{k},-} \right)\, ,
\label{Eq:Hform}
\end{align}
where the operators $A^\dagger_{m, \vec{k}, \pm}, A_{m, \vec{k}, \pm}$ are linear combinations of the form
\begin{align}
   & A_{m, \vec{k},\sigma} = \sum_{n=1}^N (u_{mn,\sigma}(\vec{k})\, b_{n,\vec{k}} + v_{mn,\sigma}(\vec{k})\, b_{n, -\vec{k}}^\dagger)\, ,
\label{Eq:mode_annihilation_flat}    
\end{align} 
obeying the usual boson commutation relations.
Here $R_+, R_-$ are non-negative integers, with $R_+ + R_- = M$, which count the number of particle and hole-like bands of the dynamical matrix respectively that have frequency $\omega > \omega_0$. The coefficients $u_{mn,\sigma}, v_{mn,\sigma}$ in (\ref{Eq:mode_annihilation_flat}) are complex numbers that can be obtained from the original system's dynamical matrix $D$ (see Sec.~\ref{Sec:Proc}), and can be arranged in $R_+ \times N$ and $R_- \times N$ dimensional matrices $u_{\sigma}, v_{\sigma}$. 
 These matrices have two important properties that we will need below. First, $u_{\sigma}, v_{\sigma}$ obey the following identities, which are consequences of the bosonic commutation relations of the operators $A_{m,\vec{k},\sigma},\, A^{\dagger}_{m,\vec{k},\sigma}$:
\begin{subequations}\label{Eq:mode_orth1}
\begin{align}
    & u_{\sigma} (\vec{k})u_{\sigma'}^{\dagger}(\vec{k})-v_{\sigma}(\vec{k})v^{\dagger}_{\sigma'}(\vec{k}) = \delta_{\sigma,\sigma'}\mathbbm{1}_{R_\sigma}\, , \label{uudagid}\\
    & u_{\sigma}(\vec{k})v^T_{\sigma'}(-\vec{k})- v_{\sigma}(\vec{k})u^T_{\sigma'}(-\vec{k}) = 0\, . \label{uvid}
\end{align}
\end{subequations}
Another important property of $u_{\sigma}, v_{\sigma}$ that we will need below is that they can always be chosen so that they depend smoothly on $\vec{k}$ -- or more precisely, they can always be chosen so that gauge-invariant combinations of $u_\sigma, v_\sigma$, such as $u_\sigma^\dagger(\vec{k}) u_{\sigma}(\vec{k})$ and $v_\sigma^\dagger(\vec{k}) v_{\sigma}(\vec{k})$ and $u_\sigma^\dagger(\vec{k}) v_{\sigma}(\vec{k})$ depend smoothly on $\vec{k}$ (see Sec.~\ref{Sec:Proc}).

%\ml{Add comment about subtlety in finding u's and v's in general non-positive definite case; see section VI}

%The choice of $u_{m,n,\sigma},\, v_{m,n,\sigma}$ notation has its origin in the BdG eigenvectors of the underlying physical system with some modification that will be clarified later in Sec.~\ref{subsec:mode}. 

%Notice that in our flattening, the dynamical matrix preserves its particle-hole symmetry, \textit{i.e.} in Fig.~\ref{Fig:Flattening} the `red' (negative-norm) bands above the shaded region are related to the negative energy `blue' (positive-norm) bands by particle-hole symmetry.
%Corresponding to the above flattened dynamical matrix, we have two equivalent choices to write the second-quantized Hamiltonian. Namely, (i) we can choose all the positive energy flat bands, (ii) or choose all the non-zero energy positive norm flat bands.
%In writing Eq.~\ref{Eq:Hform}, we choose the later form. 
%As we will see, this form allows us to interpret the second quantized Hamiltonian as sum of two spectrally separated Hamiltonians and hence the topological classification of the particular gap of interest can be viewed as sum of two topological numbers coming from the two Hamiltonian sectors. 
%This interpretation of gap topology is identified as $\mathbb{Z} \oplus \mathbb{Z}$ (or $\mathbb{Z}_2 \oplus \mathbb{Z}_2$) in Ref.~\cite{Kawabata2019}.

%\aash{Should define $N$ in words.}

%The Hamiltonian $H$ does not conserve the number of bare bosons ($\hat{b}^\dagger \hat{b}$). Since we consider the bosonic system in its dynamically stable regime, we assume that the Hamiltonian is diagonalizable, \textit{i.e.} it has a complete set of eigenmodes, which can be counted as $R_+$ modes with eigenvalue $\omega = 1$, $R_-$ modes at $\omega = -1$ and rest of $N-R_+ - R_-$ modes are at zero energy.

Once we write the resulting flattened Hamiltonian $H$ in the form (\ref{Eq:Hform}), we can now define our mapping. We map $H \rightarrow \tilde{H}$ where $\tilde{H}$ is the following particle number conserving Hamiltonian:
\begin{subequations}\label{Eq:Main_map}
\begin{align}
    & \tilde{H}  = \sum_{\vec{k}} \sum_{m=1}^N \sum_{n=1}^N Q_{mn}(\vec{k})\, b_{m,\vec{k}}^\dagger b_{n,\vec{k}}\,  ,\\   
    & Q(\vec{k}) = u_+^\dagger(\vec{k}) u_+(\vec{k}) - u_-^\dagger(\vec{k}) u_-(\vec{k})\, ,
\end{align}
\end{subequations}
This mapping amounts to simply setting the ``hole'' components $v_{mn,\sigma}(\vec{k})$ of each quasiparticle wavefunction to $0$ in Eq.~(\ref{Eq:Hform}).

The $H \rightarrow \tilde{H}$ mapping has two properties that make it a useful tool for studying the topology
of bosonic Hamiltonians. Both of these properties are proven in Sec.~\ref{Sec:interp}. The first property is that for any $H$ in the form of Eq.~(\ref{Eq:Hform}), the corresponding particle number conserving Hamiltonian $\tilde{H}$ is (i) short-ranged in real space, and (ii) the dynamical matrix of $\tilde{H}$ has a spectral gap around $\omega = \pm 1/2$. The second property is that $H$ can be continuously deformed into $\tilde{H}$ without closing the spectral gap around $\omega = \pm 1/2$. More precisely, the following theorem holds:
%\tilde{H} has $R_+$ modes with positive energy, $R_-$ modes with negative energy and $N - R_+ - R_-$ modes with zero energy.
\\
\begin{theorem}
\label{mainthm}
There exists a continuous, one parameter family of interpolating Hamiltonians, $H_{\epsilon}$ with $\epsilon\in [0,1]$, and with $H_0 = H$, and $H_1 = \tilde{H}$, such that:
\begin{enumerate}
\item $H_\epsilon$ is short-ranged in real space.
\item $H_\epsilon$ is dynamically stable.
\item The dynamical matrix of $H_\epsilon$ has a spectral gap around $\omega = \pm 1/2$.
\end{enumerate}
\end{theorem}
Theorem \ref{mainthm} has a number of implications: \\

\noindent \textbf{1.} The mapping $H \rightarrow \tilde{H}$ is ``topology preserving'':  that is, two dynamically stable bosonic Hamiltonians $H'$ and $H''$ can be continuously connected within the space of dynamically stable bosonic Hamiltonians, if and only if their number conserving partners, $\tilde{H}'$ and $\tilde{H}''$, can be continuously connected in the space of number conserving Hamiltonians. Here, when we say two Hamiltonians can be ``continuously connected'' we mean that there exists a continuous interpolation between the two Hamiltonians that preserves the spectral gap around $\omega = \pm 1/2$ (as in the statement of Theorem \ref{mainthm} above). To prove the ``if'' direction, we use the fact that $H'$ can be continuously connected to $\tilde{H}'$ and $H''$ can be continuously connected to $\tilde{H}''$. It then follows that if $\tilde{H}'$ and $\tilde{H}''$ can be continuously connected to each other, then $H'$ and $H''$ can also be continuously connected to each other. Likewise, to prove the ``only if'' direction, note that the $H\rightarrow \tilde{H}$ mapping is \emph{continuous}, and therefore any interpolation between $H'$ and $H''$ can be mapped onto a corresponding number conserving interpolation between $\tilde{H}'$ and $\tilde{H}''$. \\

    %\item{A dynamically stable anomalous bosonic Hamiltonian  can be topologically classified under an AZ class.}
    
    %\textit{Proof}. Since the map $\tilde{H}$ adiabatically connects to $H$ and since $\tilde{H}$ is number conserving, it can be classified as free fermions under enumerations of particle-hole, time-reversal and charge-conjugation symmetries.
    
    %\item The topological classification of a particular gap of $H$ can be characterized by a corresponding particle number conserving Hamiltonian $\tilde{H}$.
    %Hence a dynamically stable anomalous bosonic Hamiltonian inherits its topology from the AZ ten-fold classification of free fermions. More precisely, the AZ classification of the number conserving map $\tilde{H}$ is also the topological classification of the original anomalous Hamiltonian $H$.

\noindent \textbf{2.} Given that the mapping $H \rightarrow \tilde{H}$ is topology preserving, we can determine the topology of $H$ by studying its partner Hamiltonian $\tilde{H}$. This is useful because $\tilde{H}$ is equivalent to a number conserving fermionic Hamiltonian and therefore its topology can be analyzed using standard methods. A key point, however, is that one needs to classify the topology of $\tilde{H}$ with respect to \emph{two} band gaps, namely the band gaps at $\omega = \pm 1/2$. As a result, the classification of $\tilde{H}$ (and hence $H$) is generally determined by \emph{two} invariants $\nu_+$ and $\nu_-$, which characterize the topology of the two band gaps of $\tilde{H}$. The structure of these band invariants depend on the symmetries of the original bosonic Hamiltonian $H$. For example, in the case where we do not impose any symmetry constraints on $H$, then the corresponding Hamiltonian $\tilde{H}$ belongs to the Altland-Zirnbauer symmetry class ``$A$''; it then follows from standard classification results that the two invariants, $\nu_+$ and $\nu_-$, take values in $\mathbb{Z}$ or $\{0\}$ depending on whether the spatial dimension is even or odd. In particular, this means that the topology of the original bosonic Hamiltonian is classified by $\mathbb{Z} \times \mathbb{Z}$ for systems with even spatial dimension, and has a trivial classification for systems with odd spatial dimension. \\
%In the special case where $M=0$, the two band invariants collapse to a single invariant $\nu = \nu_+ = \nu_-$.

%The two band invariants $\nu_+$ and $\nu_-$ have a simple physical interpretation in terms of $H$: they describe the structure of the positive norm and negative norm boundary modes of $H$ within the $\omega = +1/2$ band gap.

\noindent \textbf{3.} The topology of $H$ is completely determined by $u_\sigma(\vec{k})$ -- i.e. the \emph{particle-like} half of the quasiparticle wavefunctions. To see this, note that $\tilde{H}$ is expressed entirely in terms of $u_+$ and $u_-$, and hence all topological properties of $H$ are determined by these two quantities. In particular, it is easy to check that the invariant $\nu_+$ discussed above is determined by $u_+$, and similarly the invariant $\nu_-$ is determined by $u_-$. \\

%since, the $N$-dimensional vectors $u_{m,\sigma}(\vec{k})$ are the band eigenvectors (up-to normalization) of the map $\tilde{H}$, they characterize the band topology of $\tilde{H}$ and hence $H$. Moreover, for the periodic system, well known expressions of fermionic topological invariants calculated on the band eigenstates $u_{m,\sigma}(\vec{k})$ characterizes the topology of the band $|\vec{k}_{m,\sigma}\rangle$. \\
 
\begin{figure}[!t]
%  \begin{tabular}{c c}
%    (a) \includegraphics[width=0.45\textwidth]{figures/E_DOS_1.png} & (b) \includegraphics[width=0.45\textwidth]{figures/E_DOS_2.png} \\
%    (c) \includegraphics[width=0.45\textwidth]{figures/E_DOS_3.png} & (d) \includegraphics[width=0.45\textwidth]{figures/E_DOS_4.png}
%  \end{tabular}
  \includegraphics[width=0.45\textwidth]{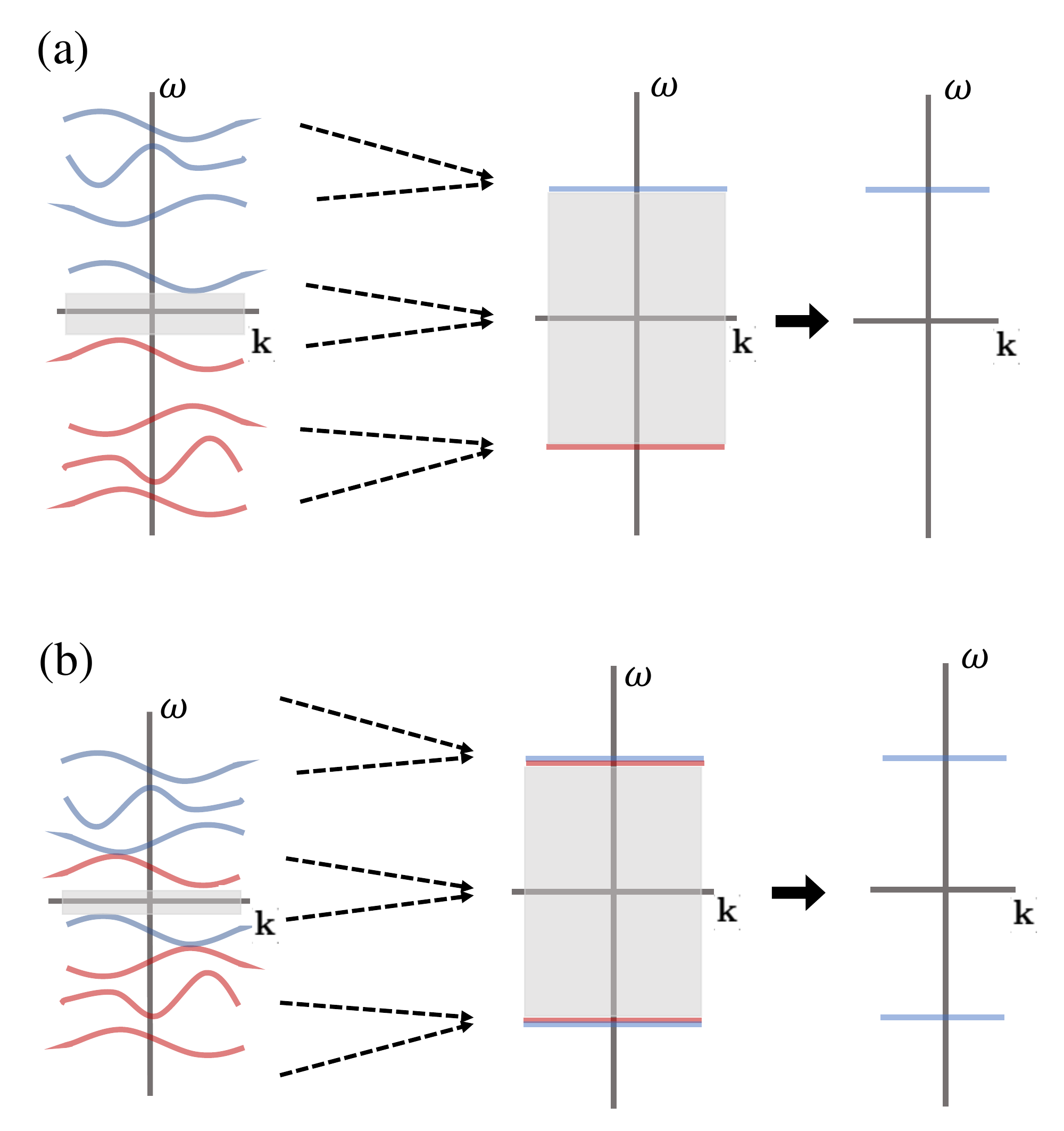}
  \caption{\label{Fig:zeroEnergy}
  Topology in the zero energy gap: (a) positive definite case, leads to a flat band system where the zero energy gap is connected to trivial vacuum, (b) relaxing the positive definite constraint can lead to situation where the zero energy gap is not directly connected to trivial state.}
\end{figure}

\noindent \textbf{4.} If $H$ is positive definite, the zero energy gap of $H$ is topologically trivial. To see this, note that if $H$ is positive definite then $R_- = 0$. If we consider a gap at zero energy, the corresponding number conserving Hamiltonian $\tilde{H}$ has $R_+ = N$ positive energy bands and no bands at zero or negative energy (see Fig.~\ref{Fig:zeroEnergy} (a)). Hence, the zero energy gap is adiabatically connected to the trivial zero energy gap of $\tilde{H}$ and must be trivial. In contrast, the zero energy gap of $H$ can be topologically non-trivial if we relax the positive definite condition (see Fig.~\ref{Fig:zeroEnergy} (b)). \\
%We discuss this point further in Sec.~\ref{subsec:TwoInvariants}. 

\noindent \textbf{5.} In the special case where $H$ is positive definite, the $H \rightarrow \tilde{H}$ mapping can be rewritten in the following simple form:
\begin{align}
    \tilde{h}= \frac{1}{4}(h + \tau_z h \tau_z + \tau_z h \tau_z h + h \tau_z h \tau_z)\, .
        \label{Qformposdef}
\end{align}
where $h = h(\vec{k})$ and $\tilde{h} = \tilde{h}(\vec{k})$ are the BdG matrices corresponding to $H$ and $\tilde{H}$ respectively:
\begin{align}
    H = \frac{1}{2}\sum_{\vec{k}} \phi^{\dagger}_{\vec{k}}\, h(\vec{k})\, \phi_{\vec{k}}\, , \quad
    \tilde{H} = \frac{1}{2}\sum_{\vec{k}} \phi^{\dagger}_{\vec{k}}\, \tilde{h}(\vec{k})\, \phi_{\vec{k}}\, ,
\end{align}
(Here, $\phi_{\vec{k}} = (b_{1,\vec{k}},\, ...,\,b_{N,\vec{k}},b^{\dagger}_{1,-\vec{k}}\, ...,\,b^{\dagger}_{N,-\vec{k}}  )^T$).
 %  \begin{align}
 %      & Q(\vec{k}) = %\frac{\mathbbm{1}+\tau_z}{4}[\tau_z\bar{h}(\vec{k})+\bar{h}(%\vec{k})\tau_z\notag\\
 %       &\hspace{1cm}+\bar{h}(\vec{k})\tau_z\bar{h}(\vec{k})%+\tau_z\bar{h}(\vec{k})\tau_z\bar{h}(\vec{k})\tau_z]\, .
 %       \label{Qformposdef}
 %   \end{align}
%We see that the final particle conserving Hamiltonian can be expressed directly in terms of $\bar{h}$, the BdG matrix corresponding to the original anomalous flat band Hamiltonian $H$ in Eq.~(\ref{Eq:Hform}).
The above expression (\ref{Qformposdef}) has a simple heuristic interpretation: the pairing terms in our original Hamiltonian $H$ generate new effective particle-conserving hopping terms in $\tilde{H}$. One can think of the pairing terms as mediating an Andreev-reflection like conversion of particles to holes, hence to second order such processes provide a new means of hopping.  Ref.~\cite{Bardyn2016} discusses such a picture explicitly for a model where it is valid to treat pairing perturbatively. We stress that Eq.~(\ref{Qformposdef}) is however an exact, non-perturbative result.
We derive Eq.~(\ref{Qformposdef}), along with its generalization to the non-positive definite case in App.~\ref{App:Hamiltonian_Map}. \\

%%%%%%%%%%%%%%%%%%%%%%%%%%%%%%%%%%%%
%%%%%%%%%%%%%%%%%%%%%%%%%%%%%%%%%%%%

\noindent \textbf{6.} It is straightforward to generalize our procedure to time reversal and crystalline symmetries. More generally, our method extends to any symmetry that does not mix boson creation and annihilation operators, i.e. any unitary or anti-unitary symmetry transformation $U$ of the form $U^{-1} b_{n,\vec{r}} U = \sum_{n', \vec{r'}} S_{nn',\vec{r} \vec{r'}} b_{n', \vec{r'}}$ for some coefficients $S_{nn', \vec{r} \vec{r'}}$. The reason that our procedure generalizes to these symmetries is the following result (which we explain below): if $H$ is invariant under the symmetry, i.e. $U^{-1} H U = H$, then $\tilde{H}$ is also invariant under the symmetry, i.e. $U^{-1} \tilde{H} U = \tilde{H}$, and furthermore the entire interpolation $H_\epsilon$ (defined in Eq.~(\ref{Eq:HInter}) below) also obeys $U^{-1} H_\epsilon U = H_\epsilon$. As a result, all of our arguments go through in the presence of the symmetry $U$. In particular, the topology of $H$ can still be determined by considering the topology of its partner Hamiltonian $\tilde{H}$ -- the only difference being that one needs to analyze the topology of $\tilde{H}$ in the presence of the symmetry constraint $U$. A simple way to prove this claim is to use the formula (\ref{Qformposdef}) for $\tilde{H}$ (and its generalization in App.~\ref{App:Hamiltonian_Map}). It is clear from this formula that the matrix $Q$ inherits all the symmetries of the BdG matrix $h$ as long as those symmetries commute with $\tau_z$, or equivalently, as long as those symmetries do not mix boson creation and annihilation operators. We conclude that $U^{-1} H U = H$ implies $U^{-1} \tilde{H} U = H$. A similar argument shows that $U^{-1} H_\epsilon U = H_\epsilon$ holds as well. \\

%In the next section, we prove Theorem \ref{mainthm}. Readers interested in the practical aspects of calculating the topological invariants can skip the next section and follow Sec.~\ref{Sec:Proc}. 

\section{Interpolation\label{Sec:interp}}

In this section, we prove Theorem \ref{mainthm}: that is, we prove the existence of an interpolating Hamiltonian $H_\epsilon$ with all the required properties (1-3). Note that once we establish this result, the other claims that we made about $\tilde{H}$ come for free, namely that (i) $\tilde{H}$ is short-ranged in real space, and (ii) the dynamical matrix of $\tilde{H}$ has a spectral gap around $\omega = \pm 1/2$. Indeed, both of these properties follow from Theorem \ref{mainthm} by considering the special case $\epsilon = 1$.

%\aash{Theorem 2 is now a claim, but there are still references to theorem 2 in the text... read carefully and fix these.}

% \aash{Text in this section needs to be updated to reflect new restatement of Theorem 1.  Seems to incorrectly mix references to Thm1, 2 and 3.  I've tried to fix, but someone else should also check carefully.}

We prove Theorem \ref{mainthm} by an explicit construction. Consider the following interpolating Hamiltonian:
\begin{align}
& H_{\epsilon} = \sum_{\vec{k}} \biggl ( \sum_{m=1}^{R_+} A_{m,\vec{k},+}^\dagger(\epsilon) A_{m,\vec{k},+}(\epsilon) \notag\\ &\hspace{2cm}- \sum_{m=1}^{R_-} A_{m,\vec{k},-}^\dagger(\epsilon) A_{m,\vec{k},-}(\epsilon) \biggr )\, ,
\label{Eq:HInter}
\end{align}
where
\begin{align}
   & A_{m, \vec{k},\sigma} (\epsilon) = \sum_{n=1}^N (u_{mn,\sigma}(\vec{k},\epsilon)\, b_{n,\vec{k}} + v_{mn,\sigma}(\vec{k},\epsilon)\, b_{n, -\vec{k}}^\dagger)\, ,
\label{Eq:mode_annihilation_inter}    
\end{align} 
and the coefficients $u_{mn,\sigma}(\vec{k},\epsilon)$, $v_{mn,\sigma}(\vec{k},\epsilon)$ are defined by
\begin{align}
    & u_{mn,\sigma}(\vec{k}, \epsilon) = u_{mn,\sigma}(\vec{k}) \, ,\quad
    v_{mn,\sigma}(\vec{k},\epsilon) = (1-\epsilon) v_{mn,\sigma}(\vec{k})\, . 
    \label{uvepsdef}
\end{align} 
By construction $H_0 = H$ and $H_1 = \tilde{H}$, and hence for $\epsilon\in[0,\,1]$, $H_{\epsilon}$ is an interpolation that connects the original, dynamically stable Hamiltonian $H$ to the particle number conserving Hamiltonian $\tilde{H}$. Notice that this interpolation is simply obtained by scaling down the `\textit{hole}' components $v_{\sigma}(\vec{k})$ of the quasiparticle wavefunctions. 

%\begin{claim} 
%\label{claim_inter}
%The interpolation Hamiltonian $H_{\epsilon}$ is (i) diagonalizable and (ii) does not close the energy gap between top %$M$ bands and the rest $2N-M$ bands for any $\epsilon$ in its domain.
%\end{claim}

%\textit{Proof}- First, we note 

We claim that $H_\epsilon$ has all the properties (1-3) listed in Theorem \ref{mainthm}. We start by proving property 2, i.e. $H_\epsilon$ is dynamically stable. The first step is to work out the commutation relations for $A_{m, \vec{k},\sigma} (\epsilon)$, $A_{m, \vec{k},\sigma}^\dagger(\epsilon)$. A straightforward calculation shows
\begin{subequations}
\begin{align}
    & [A_{m, \vec{k}, \sigma}(\epsilon),\, A_{m', \vec{k}', \sigma'}(\epsilon)] = 0 \, , \label{aacomm0} \\
    & [A_{m, \vec{k}, \sigma}(\epsilon),\, A_{m', \vec{k}', \sigma'}^\dagger(\epsilon)] = \alpha^{\sigma\sigma'}_{mm'}(\vec{k},\epsilon) \delta_{\vec{k},\vec{k}'}\, , \label{aadagcomm}
\end{align}
\end{subequations}
where the coefficient $\alpha^{\sigma\sigma'}_{mm'}(\vec{k},\epsilon)$ is an element of the $R_{\sigma}\times R_{\sigma'}$-dimensional matrix
\begin{align}\label{Eq:alpha_mat}
    & \alpha^{\sigma\sigma'}(\vec{k},\epsilon) = u_{\sigma}(\vec{k},\epsilon)u^{\dagger}_{\sigma'}(\vec{k}, \epsilon)-v_{\sigma}(\vec{k},\epsilon)v^{\dagger}_{\sigma'}(\vec{k},\epsilon)\, .
\end{align}
Here Eq.~(\ref{aacomm0}) follows from the identity (\ref{uvid}) together with the definition (\ref{uvepsdef}).

Next, using the above commutation relations, we derive
\begin{align}
&[A_{m, \vec{k},+}(\epsilon),\, H_{\epsilon}] = \sum_{m'=1}^{R_{+}} \alpha_{mm'}^{++}(\vec{k}, \epsilon) A_{m', \vec{k},+}(\epsilon) \nonumber \\
&\hspace{3cm} - \sum_{m'=1}^{R_{-}} \alpha_{mm'}^{+-}(\vec{k}, \epsilon) A_{m', \vec{k},-}(\epsilon) \nonumber \\
&[A_{m, \vec{k},-}(\epsilon),\, H_{\epsilon}] = \sum_{m'=1}^{R_{+}} \alpha_{mm'}^{-+}(\vec{k}, \epsilon) A_{m', \vec{k},+}(\epsilon) \nonumber \\
&\hspace{3cm} - \sum_{m'=1}^{R_{-}} \alpha_{mm'}^{--}(\vec{k}, \epsilon) A_{m', \vec{k},-}(\epsilon)
\label{Heiseq}
\end{align}
%\begin{align}
%    & [\hat{A}_{m, \vec{k},\sigma}(\epsilon),\, H_{\epsilon}] = \sum_{m'=1}^{R_{}} \alpha_{mm'}^{\sigma,\sigma} \hat{A}_{m', \vec{k},\sigma}(\epsilon)\, \notag\\
%    &\hspace{3cm} - \sigma \sum_{m'=1}^{R_{-\sigma}} \alpha_{mm'}^{\sigma,-\sigma} \hat{A}_{m', \vec{k},-\sigma}(\epsilon)  \, ,
%\end{align}
%governs the spectral evolution of the interpolation.
Eq.~(\ref{Heiseq}) is useful because it determines the Heisenberg equation of motion for $A_{m, \vec{k},\sigma}(\epsilon)$, which in turn determines the normal mode spectrum of $H_\epsilon$. In particular, the frequency spectrum of $H_{\epsilon}$ can be obtained by finding the eigenvalues of the $M \times M$ matrix appearing on the right hand side of (\ref{Heiseq}), which we denote by 
\begin{align}
     & X_{\epsilon}(\vec{k}) = \begin{pmatrix} \alpha^{++}(\vec{k}, \epsilon) & -\alpha^{+-}(\vec{k}, \epsilon) \\ \alpha^{-+}(\vec{k}, \epsilon) & -\alpha^{--}(\vec{k}, \epsilon) 
\end{pmatrix}\, ,
\end{align}
(To be precise, the eigenvalues of $X_\epsilon$ determine $M$ pairs of eigenvalues ($\pm \omega$) of the dynamical matrix of $H_\epsilon$; the remaining $N-M$ pairs of eigenvalues are pinned at $\omega = 0$).

Now, to prove that $H_\epsilon$ is dynamically stable, we need to show that the dynamical matrix of $H_\epsilon$ is diagonalizable and has real eigenvalues. In view of the above analysis, it is enough to show that $X_{\epsilon}(\vec{k})$ is diagonalizable and has real eigenvalues~\footnote{Strictly speaking, to make the connection between the spectrum of $X_\epsilon$ and $H_\epsilon$, we also need to check that that the $A_{m, \vec{k}, \pm}(\epsilon)$ operators are linearly independent. This linear independence follows immediately from the fact that $X_\epsilon(\vec{k})$ is a non-degenerate matrix, as shown below.}. For this purpose, it is useful to write
\begin{align}
X_{\epsilon}(\vec{k}) = Y_{\epsilon}(\vec{k})\, Z\, ,
\end{align}
where 
\begin{align}
    & Y_{\epsilon}(\vec{k})  = \begin{pmatrix} \alpha^{++}(\vec{k},\epsilon) & \alpha^{+-}(\vec{k},\epsilon) \\ 
    \alpha^{-+}(\vec{k}, \epsilon) & \alpha^{--}(\vec{k},\epsilon) 
    \end{pmatrix} , \quad \quad Z = \begin{pmatrix} \mathbbm{1}_{R_+} & 0 \\ 0 & -\mathbbm{1}_{R_-} \end{pmatrix}\, .
\end{align}
We claim that $Y_\epsilon \equiv Y_{\epsilon}(\vec{k})$ is a Hermitian matrix and all of its eigenvalues are larger than or equal to $1$. To see this, we note that $Y_{\epsilon}$ can be re-written in the form
\begin{align}
     Y_{\epsilon} &= V_\epsilon \cdot \begin{pmatrix} \mathbbm{1}_N & 0\\
    0 & -(1-\epsilon)^2\mathbbm{1}_N
    \end{pmatrix}\cdot V^{\dagger}_\epsilon \notag\\
    &= V_\epsilon \left [\begin{pmatrix} \mathbbm{1}_N & 0\\
    0 & -\mathbbm{1}_N
    \end{pmatrix} + \begin{pmatrix}
 0_N & 0 \\ 0 & \epsilon(2-\epsilon) \mathbbm{1}_N \end{pmatrix} \right] V^\dagger_\epsilon \notag\\
    &=\mathbbm{1}_{R_+ + R_-} + V_\epsilon \cdot \begin{pmatrix}
 0_N & 0 \\ 0 & \epsilon(2-\epsilon) \mathbbm{1}_N \end{pmatrix} \cdot V^\dagger_\epsilon \, ,
\label{Eq:Yexp}
\end{align}
where
\begin{align}
    & V_\epsilon(\vec{k}) = \begin{pmatrix} u_{+}(\vec{k}, \epsilon) & v_{+}(\vec{k}, \epsilon) \\ u_{-}(\vec{k}, \epsilon) & v_{-}(\vec{k}, \epsilon) \end{pmatrix}\, .
\end{align}
Here the first equality follows from Eq.~(\ref{Eq:alpha_mat}) and the third equality follows from Eq.~(\ref{uudagid}). From Eq.~(\ref{Eq:Yexp}), we can see that $Y_\epsilon$ is a sum of an identity matrix and a positive semi-definite matrix. It follows that all the eigenvalues of $Y_\epsilon$ are larger than or equal to $1$, as we wished to show.

Given that $Y_\epsilon$ is a Hermitian and has eigenvalues larger than or equal to $1$, it follows that $Y_\epsilon$ has a Hermitian square root. We can then write
\begin{align}\label{Eq:Sqrt_relation}
    & Y^{-1/2}_{\epsilon} X_{\epsilon} Y^{1/2}_{\epsilon} = Y^{1/2}_{\epsilon} Z Y^{1/2}_{\epsilon}\, .
\end{align}
The right-hand-side (RHS) above is Hermitian; therefore it is diagonalizable and has a real eigenvalue spectrum. Since $X_{\epsilon}$ is equivalent to the matrix on the RHS, via a similarity transformation, it follows that $X_{\epsilon}$ is also diagonalizable and has a real eigenvalue spectrum. This completes our proof of property 2 of Theorem~\ref{mainthm}.

%\aash{I would keep some of the intermediate steps in Michael's note that leads to Eq. 21}
Next, we show that $H_\epsilon$ satisfies property 3 of Theorem~\ref{mainthm}, i.e. the dynamical matrix of $H_\epsilon$ has a spectral gap around $\omega = \pm 1/2$. The first step is to invert Eq.~(\ref{Eq:Sqrt_relation}) to obtain
\begin{align}\label{Eq:sqrt_inv}
     & Y^{-1/2}_{\epsilon} X^{-1}_{\epsilon} Y^{1/2}_{\epsilon} = Y^{-1/2}_{\epsilon} Z Y^{-1/2}_{\epsilon}\,\, .
\end{align}
We claim that all of the eigenvalues of the matrix on the RHS of Eq.~(\ref{Eq:sqrt_inv}) have absolute value of at most $1$. To see this, notice that for any vector $x$ 
\begin{align}\label{Eq:}
\|Y^{-1/2}_{\epsilon} Z Y^{-1/2}_{\epsilon} x \| &\leq \| Z Y^{-1/2}_{\epsilon} x \| \nonumber \\
&= \|  Y^{-1/2}_{\epsilon} x \| \nonumber \\
&\leq \|  x \|
\end{align}
where $\|\cdot\|$ denotes the usual $L^2$ norm. Here, the first and third inequalities follow from the fact that all the eigenvalues of $Y^{-1/2}_{\epsilon}$ have absolute value of at most $1$, while the second inequality follows from the fact that $Z$ is unitary.

Now, since the eigenvalues of the matrix on the RHS of Eq.~(\ref{Eq:sqrt_inv}) have absolute value of at most $1$, we conclude that the same is true for $X^{-1}_{\epsilon}$. Hence, all the eigenvalues of $X_{\epsilon}$ must have absolute value of at \emph{least} $1$. Given that $X_{\epsilon}$ determines the nonzero eigenvalues of the dynamical matrix of $H_\epsilon$, it follows in particular that the dynamical matrix of $H_\epsilon$ has a spectral gap around $\omega = \pm 1/2$. This completes our proof of property 3 of Theorem~\ref{mainthm}.

Finally we now show that $H_\epsilon$ obeys property 1 of Theorem~\ref{mainthm}, i.e. $H_\epsilon$ is short-ranged in real space. To prove this, it suffices to show that 
the matrix elements of $H_\epsilon$ (in the BdG description) depend smoothly on $\vec{k}$. This is clear since the BdG matrix for $H_\epsilon$ is built out of linear combinations of gauge invariant combinations of $u$'s and $v$'s, in particular: $u_\sigma^\dagger(\vec{k}) u_{\sigma}(\vec{k})$,  $v_\sigma^\dagger(\vec{k}) v_{\sigma}(\vec{k})$, and $u_\sigma^\dagger(\vec{k}) v_{\sigma}(\vec{k})$. These gauge invariant combinations depend smoothly on $\vec{k}$, as noted in Sec.~\ref{Sec:result}. Putting this all together, this completes the proof of Theorem~\ref{mainthm}.

%Since all of these matrices are gauge invariant and hence depend smoothly on $\vec{k}$, the interpolation is smooth in $\vec{k}$-space. Since in case of topological bands, a smooth gauge choice in the entire $\vec{k}$-space is not possible for the matrices $u_{\sigma}(\vec{k})$ and $v_{\sigma}(\vec{k})$ individually, the gauge invariant method used here makes it suitable for numerical implementation. Moreover, as a result, the real space form of $H_{\epsilon}$ can be obtained by simple Fourier transform.Since $H_{\epsilon}(\vec{k})$ is smooth in $\vec{k}$-space, it is short-ranged in real space. This proves the property 3 listed in Theorem 1.
%ombining the smoothness of the interpolation, along with Claim 1 and the fact that $H_{0} = H$ and $H_{1} = \tilde{H}$, completes the proof of Theorem 1. 

In App.~\ref{app:fermions}, we discuss the fermionic analog of the the above interpolation, as well as the analog of the $H \rightarrow \tilde{H}$ map. We show that both the map and the interpolation are problematic in the fermionic case because the energy gap can close at $\epsilon = 1$.
%because both the int a similar interpolation on fermionic system leads to decrease in the energy gap and can indeed close under some circumstances.
%\aash{Where is theorem 3???}

%Since the original anomalous Hamiltonian $H$ and the number conserving Hamiltonian $\tilde{H}$ are connected adiabatically via the interpolation $H_{\epsilon}$, topologically $H$, $H_{\epsilon}$ and $\tilde{H}$ are equivalent.
%Hence, characterization of $\tilde{H}$ necessarily implies the characterization of $H$. 

%%%%%%%%%%%%%%%%%%%%%%%%%%%%%%%%%%%%%%%%%
\section{Implications for classification\label{Sec:implication}}

Having established our topology-preserving map between a general dynamically-stable bosonic Hamiltonian $H_B$ of the form (\ref{Eq:system_ham}) and a particle conserving Hamiltonian $\tilde{H}$, we now have a simple and direct means to classify the topology of the target gap $\omega = \omega_0$ of $H_B$ using the standard methods developed for (particle conserving) fermionic Hamiltonians.

\subsection{Positive definite case}

The simplest case is where the original Hamiltonian $H_B$ is positive definite. In this case $R_{-}=0$ in Eq.~(\ref{Eq:Hform}), so our final particle conserving Hamiltonian $\tilde{H}$ has a single bandgap centered at $\omega = 1/2$ (i.e.~all bands are either at $\omega=0$ or $\omega \geq 1$). Our problem reduces to analyzing the topology of this single bandgap of $\tilde{H}$. The latter problem can be addressed with standard methods since $\tilde{H}$ is equivalent to a particle conserving fermionic Hamiltonian. 
%If we are interested in bosonic systems $H_B$ with time reversal symmetry or crystalline symmetries, then we can proceed in exactly the same way, except that we now need to study particle conserving fermionic Hamiltonians with the appropriate symmetries.

A general feature of positive definite systems that is worth emphasizing is that they are characterized by a \emph{single} topological band invariant, since $\tilde{H}$ has only a single bandgap. This should be contrasted with the general case discussed below.     

%We stress that the ultimate goal of classification is to address whether two different bosonic Hamiltonians can be connected by a smooth interpolation that keeps a target bandgap open while preserving relevant symmetries.  As discussed above in Sec.~\ref{Sec:interp}, our mapping provides an explicit procedure for constructing the required interpolation (and correspondingly, for identifying cases where it is impossible to construct such an interpolation).  

%Recall that symmetry is a key aspect of the AZ scheme. Within our approach, the relevant symmetries (presence of absence of time-reversal, particle and or chiral symmetries) can be directly identified from the first quantized Hamiltonian matrix $Q(\vec{k})$ corresponding to $\tilde{H}$. 

%\aash{Do we want that silly schematic figure showing explicitly how you connect to bosonic pairing Hamiltonians via two particle conserving Hamiltonians?  Could even put this on page one to immediately make it clear what our approach is.}

%%%%%%%%%%%%%%%%%%%%%%%%%%
\subsection{General case: two invariants and the possibility of edge instabilities}
\label{subsec:TwoInvariants}

The situation becomes more interesting in the general case where the starting bosonic Hamiltonian $H_B$ is dynamically stable but not positive definite.  In this case, both $R_+$ and $R_{-}$ in Eq.~(\ref{Eq:Hform}) can be non-zero, and our final particle conserving Hamiltonian $\tilde{H}$ has two independent bandgaps centered at $\omega = \pm 1/2$ (i.e.~the system has bands at $\omega = 0$ and $|\omega| \geq 1$). Classifying the topology of the bandgap at $\omega = \omega_0$ in our original bosonic Hamiltonian now corresponds to the simultaneous classification of two bandgaps in a particle-conserving Hamiltonian. Each of these bandgaps can be characterized using standard methods developed for particle conserving fermionic systems.

The net result is that the original system's bandgap of interest is now generically described by two independent topological invariants -- one for each bandgap of $\tilde{H}$.  This has a simple physical interpretation:  with open boundary conditions, our original system could in principle support both particle-like and hole-like edge states in the bandgap of interest (centered at $\omega = \omega_0$), and the two invariants we obtain characterize each of these sets of edge states separately.

For example, consider two dimensional bosonic Hamiltonians without any symmetry constraints. In this case, the number conserving Hamiltonian $\tilde{H}$ is in the Altland-Zirnbauer class ``A'', so the relevant band invariant is the Chern number. This means that the original bosonic Hamiltonian is characterized by two Chern numbers, $\nu_+$ and $\nu_-$, where
%\begin{align}\label{Eq:twoInvaraints1}
%    & \nu = \nu_+ \oplus \nu_-\, \, ,
%\end{align}
\begin{subequations}\label{Eq:twoInvariants2}
\begin{align}
    & \nu_+ = -\frac{i}{2\pi}\sum_{n\in R^+} \int d^2\vec{k} \nabla_{\vec{k}}\times \langle \mathcal{U}_n(\vec{k}) |\nabla_{\vec{k}}| \mathcal{U}_{n}(\vec{k})\rangle\, ,\\
    & \nu_- = \frac{i}{2\pi}\sum_{n\in R^-} \int d^2\vec{k} \nabla_{\vec{k}}\times \langle \mathcal{U}_n(\vec{k})|\nabla_{\vec{k}}|\mathcal{U}_n(\vec{k})\rangle\, .
\end{align}
\end{subequations}
Here, the gap Chern numbers $\nu_{+}$ is defined by summing over the band Chern numbers of the $R_+$ bands of $\tilde{H}$ above $\omega = 1/2$. Likewise, $\nu_-$ is defined by summing over the band Chern numbers of the $R_-$ bands of $\tilde{H}$ below $\omega = -1/2$. The $| \mathcal{U}_{n}(\vec{k})\rangle$ denote the band eigenvectors of $\tilde{H}$. Equivalently, if we think about these invariants from the point of view of the original bosonic Hamiltonian, the two gap Chern numbers, $\nu_+$ and $\nu_-$, can be interpreted as characterizing (respectively) particle-like and hole-like bands sitting above the positive energy gap at $\omega = \omega_0$. 
%An alternate but equivalent view is that we are effectively characterizing two bandgaps of the original system: that at $\omega = \omega_0$ and that at $\omega = -\omega_0$. 

In the positive definite case, the invariant $\nu_-$ is always zero since $R_- = 0$. In contrast, in the general case, $\nu_+$ and $\nu_-$ can both be nonzero. These two invariants are generally independent of each other, except in the case of a zero-energy gap: $\omega_0 = 0$. In that case, $\nu_+ = \nu_-$, since there is no zero eigenvalue of $\tilde{H}$ in that case and gaps at $\omega = \pm 1/2$ are connected. The possibility of two distinct invariants for the general bosonic problem was noted in Ref.~\cite{Kawabata2019}, though discussed somewhat differently.    

Perhaps our most crucial observation here is that the existence of two non-zero invariants in a specific gap of our original dynamical matrix can have profound observable consequences.  In this situation, the open boundary condition version of our system is guaranteed to have both particle-like and hole-like edge states in this target gap.  This opens up the possibility of having a system whose dynamical stability is sensitive to boundary conditions.  While the periodic boundary condition system is fully dynamically stable, the coexistence of particle and hole-like edge states at the same frequency means that (in the absence of additional selection rules) an arbitrarily small pairing perturbation will lead to dynamical instability.  Specific examples of this phenomenon have been discussed both in 1D \cite{Barnett2013} and 2D bosonic pairing Hamiltonians \cite{Barnett2015,Peano2016a}; we also provide another specific example in Sec.~\ref{Sec:num} below. In some ways, this strong sensitivity to boundary conditions can be viewed as the inverse of the recently discussed ``non-Hermitian skin effect'' (see e.g.~\cite{Yao2018,McDonaldPRX2018}), where certain non-Hermitian lattice Hamiltonians exhibit instability with periodic boundary conditions (i.e.~complex spectra), but not with open boundary conditions.

\subsection{Relation to previous work}
\label{subsec:previous}

It is helpful to put our work and results in the context of other studies also addressing topology in bosonic pairing Hamiltonians.
Chern numbers for bosonic pairing Hamiltonians were introduced in Ref.~\cite{Shindou2013}, and generalized to disordered systems in Ref.~\cite{Peano2018a}.  
As mentioned in the introduction, a comprehensive topological classification scheme for general non-Hermitian systems was recently achieved~\cite{Kawabata2019,Zhou2019}.
Although the bosonic systems considered here can be viewed
as a special case of a non-Hermitian problem \cite{Lieu2018,Kawabata2019}, our more tailored approach has several advantages.  

One advantage of our approach is the simplicity of our map, $H \rightarrow \tilde{H}$. As discussed earlier, our map simply sets the $v_{mn,\sigma}(\vec{k})$ components of each quasiparticle wave function to zero. The simplicity of this operation makes it a particularly transparent tool for analyzing bosonic topology. Also, from a practical perspective, our map does not require any momentum-space patching~\cite{Kawabata2019}, which makes it especially suitable to numerical implementation. 
%For topologically non-trivial Bloch states, often it is impossible to find a smooth global gauge choice~\cite{Thouless1982,Thonhauser2006}.A workaround this problem is to divide the $\vec{k}$-space into subregions with different  gauge choices. One then requires a procedure to patch together different subregions at the boundary~\cite{Kawabata2019}. 

%\aash{Need to explain this a bit, add a reference?}

Another advantage of our approach is that our map $H \rightarrow \tilde{H}$ is naturally suited to constructing topological band invariants for bosonic systems. To see why, consider a dynamically stable bosonic Hamiltonian that has a gap at $\omega = \omega_0$. Suppose, for simplicity, that this Hamiltonian has $M$ particle-like bands and no hole-like bands above this gap. Finding a band invariant for such a system amounts to finding a quantity $\nu$ with two properties: (i) $\nu$ characterizes the topology of the $\omega = \omega_0$ gap, and (ii) $\nu$ is expressed in terms of the spectral projector $P$ onto the $M$ bands above $\omega = \omega_0$. To find such a quantity, consider the corresponding particle conserving Hamiltonian $\tilde{H}$. This Hamiltonian has a gap at $\omega = 1/2$ and has $M$ bands above this gap. Let $\tilde{P}$ be the spectral projector onto these $M$ bands. A crucial feature of our mapping is that $\tilde{P}$ is completely determined by $P$, and is independent of any other information. Therefore, if we know how to write down a band invariant for the particle conserving projector $\tilde{P}$, then we can use our mapping to express this invariant in terms of the original spectral projector $P$. In this way, we can construct a band invariant for the original bosonic system, using well-known invariants for particle conserving systems. We note that such a construction is not possible for other mappings, such as the one described in Ref.~\cite{Kawabata2019}: in that case the particle conserving projector $\tilde{P}$ depends not just on $P$ but on the entire dynamical matrix of $H$, so there is no straightforward way to convert a band invariant for $\tilde{P}$ into a corresponding invariant for $P$.

%\new{Equally important, our map projects the system to only particle-like half of the system and requires only a small set of eigenvectors that are above a particular band gap for the purpose of calculating topological invariant.This is in contrast to Ref.~\cite{Kawabata2019}, which requires working with the full dynamical matrix. Moreover, our map to a particle number conserving Hamiltonian makes it clear that the well known fermionic topological invariants calculated on the number conserving map serve as the bulk topological invariant for the original quadratic bosoninc system.
%}

Previous studies \cite{Susstrunk2016,Lu2018,Lein2019} have also made use of alternative mappings to effective fermionic models to study band topology of positive definite bosonic pairing Hamiltonians (both at zero and non-zero energies).
Unlike our work, these approaches are limited to positive definite Hamiltonians.  There are also other important differences.  
Some of these previous works construct a similarity transformation between the original bosonic dynamical matrix $D$ of interest and a Hermitian matrix $H_f$~\cite{Susstrunk2016,Lu2018}.  The latter can be interpreted as a fermionic Hamiltonian which is isospectral to $D$.  Analogous mappings have been used even earlier to study disordered bosonic systems \cite{Gurarie2003}.  Crucially, they are not surjective:  not every fermionic Hamiltonian corresponds to a valid bosonic dynamical matrix.  As a result, they do not a priori enable classification.  

To be concrete, suppose two bosonic systems (dynamical matrices $h_{d,1}$ and $h_{d,2}$) are mapped (respectively) to fermionic Hamiltonians $H_{f,1}$ and $H_{f,2}$.  If these fermionic Hamiltonians are topologically equivalent, it is tempting to conclude that so are $h_{d,1}$,$h_{d,2}$.  However, to rigorously show this would require a family of bosonic dynamical matrices that smoothly interpolate between $h_{d,1}$ and $h_{d,2}$ (while preserving the gap of interest).  As the boson to fermion similarity transformation is not surjective, it does not guarantee the existence of such an interpolation. In contrast, the approach in our work (which is not just a simple similarity transformation) explicitly constructs the required interpolation.

An alternate approach (also restricted to the positive definite case) was recently presented in Ref.~\cite{Xu2020}.  The key idea here is that if $H_f$ is a valid, Hermitian fermionic BdG matrix, then $H_f^2$ is a valid bosonic BdG matrix.  Unlike our approach, the mapping here involves a fermionic system with pairing terms (and not a particle conserving Hamiltonian).  Further, as noted in Ref.~\cite{Xu2020}, this approach cannot be directly used for all positive definite bosonic systems, as not all such systems have BdG Hamiltonians of the requisite form.  

Finally, we note that Ref.~\cite{Flynn2020} presented an approach for unitarily mapping dynamically-stable quadratic bosonic Hamiltonians to particle-conserving models.  While this approach could also deal with positive non-definite systems, it has many crucial differences from our approach.  Unlike our mapping, the method of Ref.~\cite{Flynn2020} is not guaranteed to generate a local particle conserving Hamiltonian. In further contrast to our work, it does not correspond to a smooth interpolation, and thus 
does not guarantee a topological equivalence between the original particle non-conserving system and its number conserving partner.

\section{Additional aspects of the mapping procedure\label{Sec:Proc}}

In this section, we discuss extra details of our mapping procedure. In particular, we explain the preprocessing step that occurs before we apply the mapping $H \rightarrow \tilde{H}$. In this step, we take a general dynamically stable boson Hamiltonian described by Eq.~(\ref{Eq:system_ham}), and then flatten its bands and separate its particle-like and hole-like modes, to produce a diagonalized second-quantized Hamiltonian of the form (\ref{Eq:Hform}).

%The procedure outlined below can be taken as an algorithm to classify and characterize an arbitrary quadratic bosonic system described by Eq.~\ref{Eq:system_ham} in its dynamically stable regime.
%One important aspect of the procedure is its gauge invariance, which makes it suitable to implement numerically.

\subsection{Band flattening\label{subsec:Flat}}

%\aash{I feel like it would be helpful to present the basics of band flattening before stating the main result.  It helps establish the context and meaning of the main result.  In contrast, the mode separation stuff coulde even be moved to an appendix.   }
%\gaurav{Check the corresponding changes below}

Band topology is ultimately a wavefunction property, independent of the specific shape of individual band dispersions.  Hence, 
as is standard, our mapping procedure starts by flattening the energy spectrum of our initial bosonic Hamiltonian (\ref{Eq:system_ham}) around the gap of interest, while keeping the wavefunction information intact. 
%This leads to Eq.~(\ref{Eq:Hform}) as the starting point of our main result, which is a flat band quadratic bosonic Hamiltonian with pairing terms. 
Here, we discuss this flattening step more explicitly.
% with dispersive  bands corresponding to the original system Hamiltonian $H_B$. 

As in Sec.~\ref{Sec:result}, we focus on particular energy gap of the dynamical matrix $D(\vec{k})$ around some frequency $\pm \omega_0$. We assume that there are $M$ bands with frequencies $\omega > \omega_0$, and $M$ bands with frequencies $\omega < - \omega_0$, and $2(N-M)$ bands with frequencies $|\omega| < \omega_0$.
% We are interested in a particular energy gap, such that counting from the top, $M$ bands lie above the energy gap of interest.
%Since the system is diagonalizable, the dynamical matrix can be represented as
%\begin{align}
%     & \hat{h}_d(\vec{k}) = S(\vec{k})\, \textup{diag} [\omega_1(\vec{k}),%\,...,\,\omega_{2N}(\vec{k})]\,S^{-1}(\vec{k})\, ,
%\end{align}
%where
%\begin{align}
%     & S(\vec{k}) = (\,|\vec{k}_1\rangle,\, |\vec{k}_2\rangle,\,...,\,|\vec{k}%_{2N}\rangle \,  )\, ,
%\end{align}
%is pseudo-unitary, because it follows the identity: $R^{-1}(\vec{k}) = S^{\dagger}(\vec{k})\,\hat{\tau}_z$.
%Here we count the states from the top of the excitation spectrum.
To flatten the dynamical matrix spectrum, we push the $M$ bands with $\omega > \omega_0$ to $\omega = 1$, and the $M$ bands with $\omega < -\omega_0$ to $\omega = -1$. We take the remaining $2(N-M)$ bands to $\omega = 0$ (Fig.~\ref{Fig:Flattening}).

In practice, this flattening step is accomplished by diagonalizing $D(\vec{k})$, and writing it in the form $D(\vec{k}) = S \Lambda S^{-1}$, where $\Lambda$ is a diagonal matrix of the form $\Lambda = \mathrm{diag}(\omega_1,... \omega_{2N})$ with $\omega_1 \geq \omega_2 ... \geq \omega_{2N}$ and $S$ is a matrix whose columns are the eigenvectors of $D$. We then define two spectral projection matrices $P_+(\vec{k})$ and $P_-(\vec{k})$ by
\begin{subequations}
\begin{align}
    P_+(\vec{k}) &= S \cdot \begin{pmatrix} \mathbbm{1}_M & 0 \\ 0 & 0_{2N-M} \end{pmatrix} \cdot S^{-1}\, ,\\
    P_-(\vec{k}) &= S \cdot \begin{pmatrix} 0_{2N-M} & 0 \\ 0 & \mathbbm{1}_M \end{pmatrix} \cdot S^{-1}\, 
\end{align}
\end{subequations}
By construction, $P_+(\vec{k})$ projects onto the eigenvectors of $D(\vec{k})$ with eigenvalues $\omega > \omega_0$, while $P_-(\vec{k})$ projects onto the eigenvectors with eigenvalues $\omega < - \omega_0$. Note that the projectors $P_{\pm}(\vec{k})$ are non-Hermitian, just like $D(\vec{k})$.

%constructing appropriate spectral projection matrices. Specifically, we define 
%\begin{subequations}
%\begin{align}
%    & P_+(\vec{k}) = \sum_{\omega_m(\vec{k}) > \omega_0} |\vec{k}_{m}\,\rangle \langle\, \vec{k}_m | \tau_z\, ,\\
%    & P_-(\vec{k}) = \sum_{\omega_m(\vec{k}) < - \omega_0} |\vec{k}_{m}\,\rangle \langle\, \vec{k}_m | \tau_z\, ,
%\end{align}
%\end{subequations}
%Here $P_+(\vec{k})$ projects onto all eigenvectors $|\vec{k}_m\>$ of $h_d(\vec{k})$ with eigenvalues $\omega_m(\vec{k}) > \omega_0$. Likewise, $P_-(\vec{k})$ projects onto all eigenvectors $|\vec{k}_m\>$ with eigenvalues $\omega_m(\vec{k}) < \omega_0$.
%Note that the projectors $P_{\pm}(\vec{k})$ are non-Hermitian.

The resulting flat band dynamical matrix, $D^{\mathrm{flat}}(\vec{k})$, is then given by
\begin{align}
    & D^{\mathrm{flat}}(\vec{k}) = P_+(\vec{k}) - P_-(\vec{k})\, .
\end{align}

%We will use these to separate the positive and negative norm modes.
%Importantly, the flattened dynamical matrix is smooth in the entire Brillouin zone and can be obtained in gauge-invariant way from the numerically obtained eigenvectors.

%\begin{align}
%    & \hat{\bar{h}}_d(\vec{k}) = S(\vec{k})\begin{pmatrix}
%    \mathbbm{1}_M & & \\
%    & \mathbb{0}_{2N-2M} &  \\
%    & & -\mathbbm{1}_{M}\\
%    \end{pmatrix}
%    S^{\dagger}(\vec{k})\,\hat{\tau}_z\, .
%\end{align}
%Moreover, the flattened dynamical matrix preserves the particle-hole symmetry.
%Subsequently, the flattened dynamical matrix can be represented in the %projection operator form

\subsection{Mode separation\label{subsec:mode}}

In addition to band flattening, the other preprocessing step that we need to explain is how to separate out the particle-like and hole-like modes of the dynamical matrix $D(\vec{k})$. If the BdG matrix $h_B(\vec{k})$ is positive definite, this mode separation problem is trivial since there are only particle-like modes at positive frequencies, i.e. $R_- = 0$.
% Hence, the space of all the positive and negative symplectic norm are well separated and this step of the procedure can be skipped in the numerical implementation.
In the non-positive definite case, the situation remains simple if at each $\vec{k}$ the dynamical matrix $D(\vec{k})$ has no degeneracies between particle-like and hole-like modes. One can then identify the particle-like and hole-like modes using the non-degenerate particle-like and hole-like bands of $D(\vec{k})$. 

The only problematic case is when $D(\vec{k})$ has a degeneracy between a particle-like state
$|\vec{k}_p\rangle$ and a hole-like state $|\vec{k}_h\rangle$ for some $\vec{k}$.  In this case arbitrary linear combinations
\begin{align}
    & |\vec{k}\rangle = \alpha |\vec{k}_p\rangle + \beta |\vec{k}_h\rangle\, ,
\end{align}
are also eigenstates of the dynamical matrix with the same eigenvalue. There is thus no way to uniquely identify a particle-like and hole-like state at these degeneracy points.

In order to deal with this case, we now describe a general procedure for separating the particle-like and hole-like modes of $D(\vec{k})$. Importantly, our procedure guarantees that the particle-like and hole-like subspaces depend \emph{smoothly} on $\vec{k}$. This smoothness is important for our procedure because it ensures that our number conserving Hamiltonian $\tilde{H}$ is short-ranged in real space.
%While our procedure is non-unique, this has no bearing on our being able to employ it in our mapping procedure.  

To begin, consider the $M$ modes of our dynamical matrix, $D(\vec{k})$, with $\omega > \omega_0$.
The corresponding eigenvectors $|\vec{k}_{m}\rangle$ form an $M$-dimensional vector space $W_+$. Our mode separation procedure is based on the following observation about $W_+$:

\begin{claim}
\label{claim_sep}
The vector space $W_+$, at any point in $\vec{k}$-space can be uniquely decomposed as a direct sum 
\begin{align}
     & W_+ = W_{++} \oplus W_{+-}\, ,
\label{Eq:decom}     
\end{align}
where $W_{++}$ and $W_{+-}$ are respectively $R_+$ and $R_-$-dimensional vector spaces, such that for any nonzero vectors $w_+\in W_{++}$ and $w_-\in W_{+-}$,
\begin{subequations}
\begin{align}
    & \textup{sgn} \,\langle w_\sigma |\,\tau_z\, | w_\sigma \rangle  = \sigma \, ,\label{Eq:sep1} \\
    & \langle w_+ |\,\tau_z\, | w_- \rangle = 0\, ,\label{Eq:sep2}\\
    & \langle w_+ \, | \, w_- \rangle = 0\, .\label{Eq:sep3}
\end{align}
\end{subequations}
\end{claim}
To understand the significance of this result, note that Eq.~(\ref{Eq:sep1}) implies that $W_{++}$ and $W_{+-}$ are particle-like and hole-like subspaces, respectively, while Eq.~(\ref{Eq:sep2}) implies that these subspaces are orthogonal with respect to the symplectic inner product. Thus Eq.~(\ref{Eq:sep1}-\ref{Eq:sep2}) guarantee that (\ref{Eq:decom}) is a valid decomposition of $W_+$ into particle-like and hole-like subspaces. On the other hand, the orthogonality condition in Eq.~(\ref{Eq:sep3}) is simply a convenient gauge choice that makes the decomposition of $W_+$ unique. 

We defer the proof of Claim \ref{claim_sep} to App.~\ref{app:mode_separation}. Here we focus on the more practical question of how to construct the two orthogonal subspaces $W_{++}$ and $W_{+-}$. The first step is to construct a Hermitian projector $\bar{P}_+$ that projects onto the same subspace as $P_+$  (i.e.~that projects on $W_+$). One way to define $\bar{P}_+$ is 
\begin{align}
    \bar{P}_+ = &\text{ Projection onto subspace of positive-eigenvalue} \nonumber \\
    &\text{ eigenvectors of $P_+ P_+^\dagger$}
\end{align}
By construction, the projector $\bar{P}_+$ has exactly $M$ eigenvectors with eigenvalue $+1$ and $2N-M$ eigenvectors with eigenvalue $0$. We then define the particle-like and hole-like subspaces $W_{++}$ and $W_{+-}$ by taking the eigenvectors of the (Hermitian) matrix $\bar{P}_+\tau_z \bar{P}_+$ as the basis vectors, and defining
\begin{align}
    & W_{++} = \text{ Subspace of positive-eigenvalue eigenvectors} \nonumber \\
    & W_{+-} = \text{ Subspace of negative-eigenvalue eigenvectors}
\label{Wpmdef}
\end{align}

In App.~\ref{app:mode_separation}, we show that the subspaces defined using the above procedure satisfy the required properties in Eqs.~(\ref{Eq:sep1})-(\ref{Eq:sep3}).
Moreover, since the projection matrix $P_+(\vec{k})$ is gauge invariant and depends smoothly on $\vec{k}$, it follows that $\bar{P}_+(\vec{k})$, $W_{++}$ and $W_{+-}$ also depend smoothly on $\vec{k}$.
The net result is a smooth method of separating particle-like and hole-like subspaces that is easy to implement numerically.  In particular, it does not require a complicated patching-together of different momentum-space patches.  Such patching procedures can often be an issue when  working with the numerically obtained wavefunctions of topological bands \cite{Kawabata2019}.
%\aash{Should we cite an example of a patching procedure to make it clear this is an issue with other approaches?  Ueda?}

Having performed the mode separation, the last step is to construct the corresponding annihilation operators $A_{m,\vec{k},\pm}$. To do this, one needs to choose an orthonormal basis of $W_{++}$ and $W_{+-}$ with respect to the symplectic inner product. Denoting these basis vectors by $w_{+, m}$ and $w_{-, m}$ respectively, we define annihilation operators by
\begin{subequations}
\begin{align}
   & A_{m,\vec{k},+} = \langle w_{+,m}(\vec{k})\, |\, \tau_z\, |\,\phi\rangle\, ,\\
   & A_{m,\vec{k},-} = \langle w_{-,m}(-\vec{k})\, |\,\kappa\tau_x\,\tau_z\,|\, \phi\rangle\, ,
\end{align}
\end{subequations}
These expressions can be compared to the form in Eq.~(\ref{Eq:mode_annihilation_flat}) to read off the required wavefunction coefficients $u_{n,m,\sigma},\, v_{n,m,\sigma}$. At this point, we are finished with all the preprocessing steps: we can now write down the diagonalized flat band Hamiltonian shown in Eq.~(\ref{Eq:Hform}), having started from a generic (dynamically-stable) bosonic Hamiltonian.

\section{Numerical results}
\label{Sec:num}

%\aash{Need to tweak discussion here to not give the impression that we somehow doubt previous works that used the symplectic Chern number....}

%\aash{Is there an interesting example where there is a $\mathbb{Z}_2$ invariant?}

We have shown that the topology of a bosonic pairing Hamiltonian $H$ can be analyzed by considering an equivalent number-conserving system, $\tilde{H}$.
%, the topology of a particular band gap can be characterized by mapping the Hamiltonian onto an equivalent number-conserving model. 
%fermionic system.two topological invariants which can be directly calculated by characterizing an equivalent 
%number-conserving fermionic system. 
%The two topological invariants can further be interpreted as  the band topological invariants of the particle -like (positive symplectic norm) and the hole-like (negative symplectic norm) bands.
In this section, we demonstrate our mapping procedure for a specific bosonic pairing Hamiltonian. We compute a topological invariant for this system using our mapping to a number-conserving Hamiltonian $\tilde{H}$. We then show that our invariant agrees with a previously known topological invariant (the ``symplectic Chern number'') which is directly expressed in terms of the pairing Hamiltonian $H$.
%and a new invariant, which is calculated from $\tilde{H}$. %(calculated for the mapped, particle-conserving model $\tilde{H}$).

\begin{figure}[!t]
%  \begin{tabular}{c c}
%    (a) \includegraphics[width=0.45\textwidth]{figures/E_DOS_1.png} & (b) \includegraphics[width=0.45\textwidth]{figures/E_DOS_2.png} \\
%    (c) \includegraphics[width=0.45\textwidth]{figures/E_DOS_3.png} & (d) \includegraphics[width=0.45\textwidth]{figures/E_DOS_4.png}
%  \end{tabular}
  \includegraphics[width=0.48\textwidth]{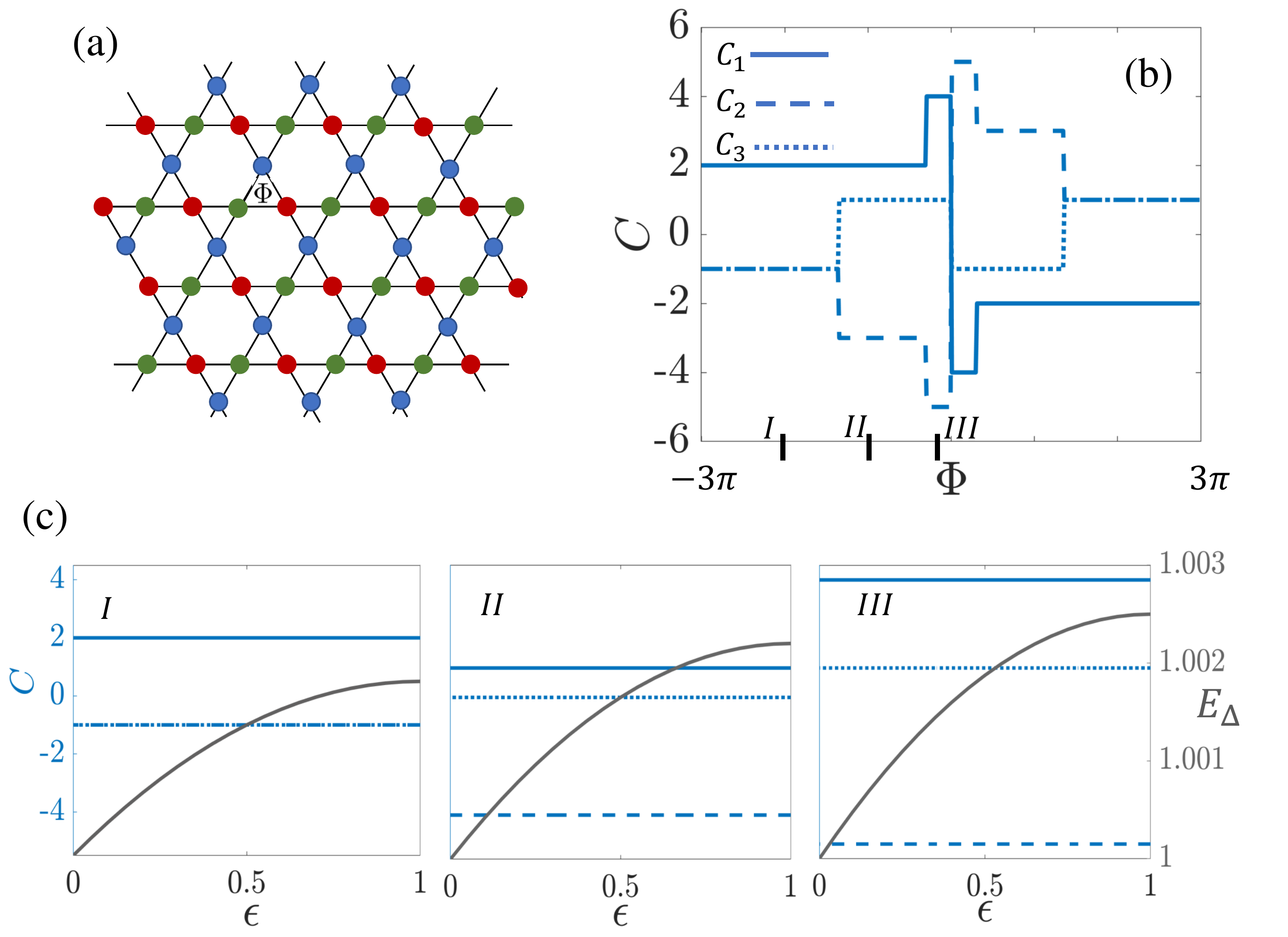}
  \caption{\label{Fig:Numerics}
  Topological phase diagram of the Kagome
  pairing model in Eq.~(\ref{eq:KagomePairingH}).
  (a) Kagome lattice, where the three sublattices $A, B, C$ are represented by blue, red and green circles respectively, and as bosons hop, they enclose flux $\Phi$ around each triangular plaquette.
  (b) The topological phase diagram for the bosonic pairing system, calculated (for each positive energy band) using our interpolated particle-conserving Hamiltonian. The calculations are done at model parameters $\omega_0 = 50$, $t = 1$, $\lambda_0 = 2.0$, $\lambda_1 = 8.0$.
  (c) For the three representative cases marked as $I, II, III$, the band Chern numbers and the gap of the interpolation Hamiltonian as the function  interpolation control parameter $\epsilon$, when the interpolation is performed for the gap between highest two bands.  We see that the Chern numbers remain invariant as expected, while the gap increases.
   }
\end{figure}

We consider a bosonic version of the model describing the quantum anomalous Hall effect on a Kagome lattice~\cite{Ohgushi2000}:
\begin{align}
    & H_0 = \sum_{\vec{j}} \omega_0 a^{\dagger}_{\vec{j}}a_{\vec{j}}
    -\sum_{\langle\vec{j},\vec{j'}\rangle} t_{\vec{j},\vec{j'}}  a^{\dagger}_{\vec{j}}a_{\vec{j'}} \, .
\end{align}
Here, a site is labeled by $\textbf{j} = (j_x, j_y, s)$, where $j_x,\,j_y\in \mathbb{Z}$ is the unit cell index and $s=A,B,C$ is the sublattice index.
The hopping term $t_{\vec{j},\vec{j'}}$ is of the form
\begin{align}
    & t_{\vec{j},\vec{j'}} = t \textup{e}^{i\varphi_{s,s'}}\, ,
\end{align}
where $\varphi_{AB} = \varphi_{BC} = \varphi_{CA} = \Phi/3$. Note that this hopping corresponds to having flux $\Phi$ through every triangular plaquette. As a result, the hopping generically breaks time reversal symmetry (the only exceptions being $\Phi = 0, \pi$). 
We further include on-site and the nearest neighbor pairing terms that break number conservation:
\begin{align}
    & H_I = - \frac{1}{2}\biggl ( \lambda_0 \sum_{\vec{j}} a^{\dagger}_{\vec{j}} a^{\dagger}_{\vec{j}} + \lambda_1 \sum_{\langle \vec{j},\vec{j'} \rangle} a^{\dagger}_{\vec{j}}a^{\dagger}_{\vec{j'}} \biggr ) + h.c. \, .
\end{align}
The Hamiltonian 
\begin{align}
    & H = H_0 + H_I\, ,
    \label{eq:KagomePairingH}
\end{align} 
was previously studied in the context of  topological phases of photonic system~\cite{Peano2015}.
In the absence of the pairing term ($\lambda_0  = \lambda_1 = 0$), band topology is the same for bosons and fermions.  In this limit, the system is in class A of the AZ scheme since the hopping term breaks $\mathcal{T}$-symmetry. 
The appropriate $\mathbb{Z}$ topological invariant is the usual Chern number and can be calculated by the TKNN formula~\cite{Thouless1982}. Counting down starting from the uppermost band, the band Chern numbers (TKNN invariants) are simply $C = [C_1,\,C_2,\,C_3] = [-\textup{sgn}(\sin\Phi/3),\, 0,\,\textup{sgn}(\sin\Phi/3)]$~\cite{Ohgushi2000}.
%\aash{Should make it clear that with the pairing terms, this Hamiltonian can remain dynamically stable without being positive definite.  Say something about parameter choices for different regimes.}

Next we introduce the pairing terms $\lambda_0,\lambda_1$.
As always, we are interested in systems that are dynamically stable.  This can be ensured by choosing a large on-site energy  $\omega_0 \gg t, \lambda_0,\lambda_1$. Such a on-site energy causes a large separation between particle-like and hole-like bands; equivalently, pairing processes are all highly detuned and hence cannot cause instability.  For large enough $\omega_0$ our system will also be positive definite, we focus on this case first.

To analyze the topology of the pairing Hamiltonian, we use our mapping procedure to construct a corresponding number-conserving Hamiltonian $\tilde{H}$. The latter Hamiltonian breaks time-reversal symmetry and therefore belongs to AZ class A. The appropriate topological invariant is again the conventional TKNN Chern number. In Fig.~\ref{Fig:Numerics}(b), we compute this Chern number invariant for the top three bands of the dynamical matrix (which are all particle-like due to positive definiteness). This calculation tells us the ``topological phase diagram'' for our original pairing Hamiltonian -- at least with respect to the top three bands.
%In Fig.~\ref{Fig:Numerics}(b) the topological phase diagram is obtained by calculating the conventional TKNN invariant for the bands of our mapped, particle-conserving Hamiltonian.  

%We will show that the system under periodic boundary condition can remain dynamically stable, even without positive definiteness.
An alternative way to analyze the topology of the pairing Hamiltonian, which was known previously, is to compute the symplectic version of the conventional TKNN invariant:
%When the pairing terms are present, 
%the bosonic BdG Hamiltonian, since it breaks $\mathcal{T}$-symmetry is in a bosonic equivalent of the AZ class D.
%an appropriate symplectic version of the conventional TKNN invariant has been used to classify the bands:
\begin{align}
    & C_n = \frac{i}{2\pi}\int d^2\vec{k} \vec{\nabla}_{\vec{k}}\times \langle \vec{k}_n |\tau_z \vec{\nabla}_{\vec{k}}| \vec{k}_n\rangle \, ,
\end{align}
This invariant is known as the ``symplectic Chern number''~\cite{Shindou2013,Peano2015,Peano2018a}.

%\aash{Discussion of classification here is a bit confusing.  Gives the impression that there was already some established way to classify these bosonic pairing Hamiltonians.  Need to make it clear that the classification follows from our procedure.  }

%As always, we are interested in systems that are dynamically stable.  This can be ensured by choosing a large on-site energy  $\omega_0 \gg t, \lambda$. Such a on-site energy causes a large separation between particle-like and hole-like bands; equivalently, pairing processes are all highly detuned and hence cannot cause instability.  For large enough $\omega_0$ our system will also be positive definite, we focus on this case first.

%In a particular gap of interest of the positive definite case, the number conserving map breaks particle-hole symmetry. Since the original bosonic model breaks $\mathcal{T}$-symmetry, the number conserving map also breaks $\mathcal{T}$-symmetry (since $\mathcal{T}$-symmetry commutes with $\tau_z$). Hence our mapping procedure will yield a number-conserving fermionic Hamiltonian belonging to AZ class A. In Fig.~\ref{Fig:Numerics}, we consider only the top three bands of the dynamical matrix, which due to positive definiteness are all particle-like. In Fig.~\ref{Fig:Numerics}(b) the topological phase diagram is obtained by calculating the conventional TKNN invariant for the bands of our mapped, particle-conserving Hamiltonian.  

To see the relationship between the symplectic Chern number and our new invariant (based on the number conserving Hamiltonian $\tilde{H}$),
we consider the $\epsilon$-dependent interpolation Hamiltonian (\ref{Eq:HInter}). Then as a function of the interpolation parameter $\epsilon$ in Fig.~\ref{Fig:Numerics}(c), we calculate the symplectic Chern number and the gap of the interpolation Hamiltonian (when the interpolation is performed for the gap between highest two bands) for three representative parameters of the model. %\ml{Need to explain which gap we are plotting. Maybe also in the caption?}. 
We can draw several conclusions from this calculation. First, we can see that the gap remains open and the symplectic Chern number is constant throughout the interpolation, as expected. Also, since the symplectic Chern number agrees with the conventional Chern number for the number conserving model at the end of the interpolation ($\epsilon = 1$), we conclude that the symplectic Chern number must also agree with our invariant for the original pairing Hamiltonian ($\epsilon = 0$).

The above results illustrate one of the main points of this paper: we can characterize the topology of our bosonic pairing Hamiltonian by mapping the system onto a particle-conserving Hamiltonian, and then calculating standard topological invariants on this mapped Hamiltonian.
In general, identifying the correct band topological invariant to calculate is a challenging problem. The advantage of our approach is that, for the case of particle-conserving Hamiltonians, many topological invariants are already known from previous work on fermionic systems. %~\cite{Bernevig2015}. \ml{consider removing this reference?} 
%Our main result allows us to leverage this knowledge in the context of bosonic pairing Hamiltonians.
%shows that the same fermionic invariants calculated for the number conserving map characterizes the band topology of interest of the original bosonic pairing Hamiltonian.
%However, the analogue of the fermionic topological invariant to the bosonic case that is also valid for the original pairing Hamiltonian is  unclear.

\begin{figure}[!htb]
%  \begin{tabular}{c c}
%    (a) \includegraphics[width=0.45\textwidth]{figures/E_DOS_1.png} & (b) \includegraphics[width=0.45\textwidth]{figures/E_DOS_2.png} \\
%    (c) \includegraphics[width=0.45\textwidth]{figures/E_DOS_3.png} & (d) \includegraphics[width=0.45\textwidth]{figures/E_DOS_4.png}
%  \end{tabular}
  \includegraphics[width=0.45\textwidth]{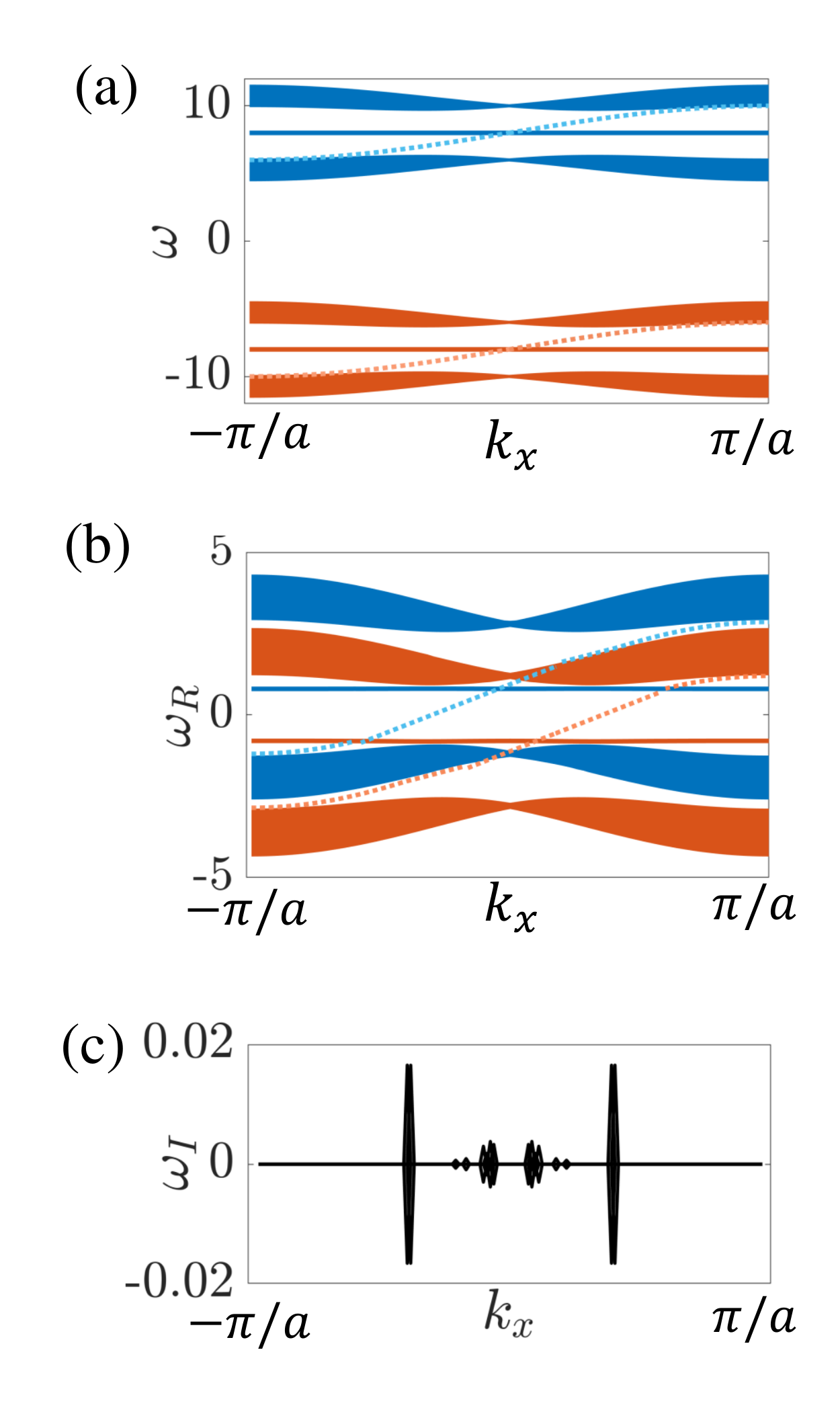}
  \caption{\label{Fig:stripeCalc}
  Spectrum of a stripe configuration of the Kagome pairing model in Eq.~(\ref{eq:KagomePairingH}). We have periodic boundary conditions in the $x$-direction and open boundary condition in the $y$-direction. 
  (a) positive definite case, $\omega_0 = 8$, $t = 1$, $\lambda_0 = 0$, $\lambda_1 = 0.05$, the particle and hole like bands are well separated. (b) Real part of the spectrum $\omega_R$, non-positive definite case, $\omega = 0.8$, $t= 1$, $\lambda_0 = 0$, $\lambda_1 = 0.05$.  Note that in the zero energy gap,  we have topological edge states from both particle and hole sectors. (c) Imaginary part $\omega_I$ of the spectrum for the same parameters as (b). Note that for both these parameter sets, the spectrum under periodic boundary conditions is purely real.  For clarity edge state at only one edge is shown. 
   }
\end{figure}

%\aash{This paragraph needs work!  Way too long....}
Next, we consider the above system in its dynamically stable and non-positive definite regime. Our goal is to present an example in which there are two nonzero band invariants in a particular band gap: i.e. the total Chern number of all the particle-like bands above the band gap is nonzero, as is the total Chern number of the hole-like bands. 

To engineer the coexistence of bulk dynamical stability and non-positive definiteness, we first take the $\lambda\rightarrow 0$ limit and tune $\omega_0$, such that some of the particle-like bands cross the hole-like bands.
%as well as the bulk bands do not have any degeneracy between particle and hole-like bands. 
Next, we introduce a weak pairing $\lambda_1$, such that spectrum under periodic boundary condition is still dynamically stable.

In Fig.~\ref{Fig:stripeCalc}, we show the spectrum of a Kagome stripe, where we have chosen periodic boundary condition in $x$-direction and open boundary condition in $y$-direction. 
%We engineer the situation similar to Ref.~\cite{Peano2016} for the Kagome model discussed here to compare the positive-deifnite and positive non-definite cases. 
In the positive definite case shown in Fig.~\ref{Fig:stripeCalc} (a) the particle-like  sector and hole-like sector are well separated, and in any particular gap only one kind of edge state (and corresponding one type of non-zero Chern number) is present, \textit{i.e.} only one sector contributes to Chern number of the periodic Hamiltonian.
The band Chern numbers counting from the top are $C = [1,0,-1,1,0,-1]$, where the first three are particle-like and the later three are hole like bands.
%In each gap, because of the complete spectral separation of particle and hole like bands there is only one type of topological invariant that is non-trivial.
In Fig.~\ref{Fig:stripeCalc} (b), we show the spectrum of the finite stripe in the non-positive definite regime. The  zero energy gap has both particle-like and hole-like edge states.
In the presence of non-zero pairing $\lambda_1$, we thus generically expect dynamical instability (as the pairing terms can resonantly populate these edge states).  This is indeed what we find numerically when considering the spectrum of a system with open boundary conditions in the $y$-direction, see Fig.~\ref{Fig:stripeCalc} (c).
Interestingly, while an infinitesimal $\lambda_1$ makes the strip system dynamically unstable, the system under periodic boundary conditions remains stable for small $\lambda_1$.  The band Chern numbers can thus be calculated using our method. 
The band Chern numbers counting from the top are $C = [1, 1,0,0,-1,-1]$.
As discussed, these Chern numbers correspond to two distinct invariants $\nu_\pm$ for each gap (one corresponding to partile-like states, the other to hole-like states).   
As an example, for the gap around zero energy, the gap Chern number $\nu_+ = \nu_- = -1$, which is consistent with the particle and hole like edge states in the finite size spectrum.

%\aash{Perhaps stating explicitly that our approach says this Kagome pairing model is in class A, and not some strange bosonic version of class D?  Or should we avoid getting into this?}

%%%%%%%%%%%%%%%%%%%%%%%%%%%%%%%%%%%%%%%%%
\section{Discussion and Conclusion\label{Sec:dis}}

In conclusion, we have presented a topology preserving map that connects an arbitrary dynamically stable quadratic bosonic system to a number conserving system.
%We also provided a procedure to obtain the map in a gauge-invariant fashion.
Our mapping allows one to analyze the topology of a gap of interest in the energy spectrum of a bosonic system by studying a corresponding gap in its number conserving partner. 
For practical purposes, our map simplifies the bosonic topological classification and characterization problem, since one only needs to consider the symmetries of the equivalent number conserving Hamiltonian and use the well known expressions for number conserving (fermionic) topological invariants.

Our explicit map has number of advantages over previous related works~\cite{Lu2018,Susstrunk2016,Kawabata2019,Xu2020,Flynn2020}. 
We explicitly present a continuous and local interpolation between space of bosonic pairing Hamiltonians and number conserving Hamiltonian and show that the relevant gaps remain open throughout the interpolation.  
Further our approach is applicable to pairing Hamiltonians that are dynamically stable but not necessarily positive definite. 

An interesting class of systems that deserve further study are bosonic systems that have a dynamically stable bulk together with unstable boundary modes. In these systems, we can unambiguously calculate bulk topological invariants using our number conserving map. At the same time, the nature of the boundary modes is not entirely clear due to their dynamic instabilities. 
We leave a more complete study of bulk-boundary correspondence for such systems for future work.

\section*{Acknowledgements}
This material is based upon work supported by the Air Force Office of Scientific Research under award number FA9550-19-1-0362, and was partially supported by the University of Chicago Materials Research Science and Engineering Center, which is funded by the National Science Foundation under 
Grant No.~DMR-1420709.  M.~L.~was supported in part by the Simons Collaboration on Ultra-Quantum Matter, which is a grant from the Simons Foundation (651440).  A.~A.~C.~acknowledges support from the Simons Foundation through a Simons Investigator award.

%%%%%%%%%%%%%%%%%%%%%%%%%%%%%%%%%%%%%%%%%%%%%%%%%%%%%%%%%%%%%%%%%%%%%%%%%%%%%%%%%%%%%%%%%%%%%%%%%%%%
\appendix
%%%%%%%%%%%%%%%%%%%%%%%%%%%%%%%%%%%%%%%%%%%%%%%%%%%%%%%%%%%%%%%%%%%%%%%%%%%%%%%%%%%%%%%%%%%%%%%%%%%%

%%%%%%%%%%%%%%%%%%%%%%%%%%%%%%%%%%%%%%%%%%%%%%%%%%%%%%%%%%%%%%%%%%%%%%%%%%%%%%%%%%%%%%%%%%%%%%%%%%%%

\section{Proof of mode separation\label{app:mode_separation}}
In this appendix, we prove that the mode separation procedure outlined in Sec.~\ref{subsec:mode}, satisfies the properties listed in Eqs.~(\ref{Eq:sep1}-\ref{Eq:sep3}).
Let $w_+ \in W_{++}$, and $w_- \in W_{+-}$ be nonzero vectors. Then $w_+$ and $w_-$ are linear combinations of eigenvectors of $\bar{P}_+ \tau_z \bar{P}_+$ with positive and negative eigenvalues, respectively.    
Therefore
\begin{align}\label{Eq:sep_prop1}
    \langle w_+ |\tau_z| w_+ \rangle &= \langle w_+ | \bar{P}_+ \tau_z \bar{P}_+ |w_+\rangle  \notag\\
                                   &\geq \lambda_{min,+} || w_+ ||^2  \notag \\
                                   &> 0\, ,
\end{align}
where $\lambda_{min,+}$ is the minimum positive eigenvalue of $\bar{P}_+ \tau_z \bar{P}_+$. (Here, the first equality follows from the fact that $\bar{P}_+ |w_+\rangle = |w_+\rangle$).
Similarly,
\begin{align}\label{Eq:sep_prop2}
     \langle w_- | \tau_z | w_- \rangle &= \langle w_- | \bar{P}_+ \tau_z \bar{P}_+ |w_- \rangle \notag \\
    &\leq\lambda_{max,-}  || w_-||^2 \notag \\
    &< 0\, ,
\end{align}
where $\lambda_{max,-}$ is the maximum negative eigenvalue of $\bar{P}_+ \tau_z \bar{P}_+$.
The above two relations combined prove Eq.~(\ref{Eq:sep1}).

Next, we note that $\bar{P}_+ \tau_z \bar{P}_+$ is Hermitian and therefore its eigenspaces are orthogonal under the usual inner product. It follows that
\begin{align}
    &\langle w_+ |  w_-\rangle = 0\, ,
\end{align}
which proves Eq.~(\ref{Eq:sep3}).

To prove Eq.~(\ref{Eq:sep2}), we note that
\begin{align}
    & \langle w_+ | \tau_z | w_- \rangle =  \langle w_+ | \bar{P}_+ \tau_z \bar{P}_+ | w_- \rangle = 0\, ,
\end{align}
where the last equality follows from the fact that $w_+ \in W_{++}$ and $\bar{P}_+ \tau_z \bar{P}_+ w_- \in W_{+-}$,  together with property in Eq.~(\ref{Eq:sep2}).

Finally, we just need to show that $W_{++}$ and $W_{+-}$ obey Eq.~(\ref{Eq:decom}) -- \textit{i.e.} these subspaces span all of $W_+$. To prove this, we need to show that $\bar{P}_+ \tau_z \bar{P}_+$ does not have any eigenvectors with zero eigenvalue that belong to the vector space $W_+$. We prove this by contradiction. Suppose that $x \in W_+$ is a nonzero vector with 
\begin{align}
    & \bar{P}_+ \tau_z \bar{P}_+ |x \rangle  = 0\, .
\end{align}
Then, for any $y \in W_+ $, we have
\begin{align}
    & \langle y | \tau_z |x \rangle =  \langle y | \bar{P}_+ \tau_z \bar{P}_+ | x\rangle = 0\, .
\label{wv1}
\end{align}
In fact, Eq.~(\ref{wv1}) also holds for any $y \in W_0$ or $y \in W_-$ (where $W_-$ denotes the eigenspace of $D(\vec{k})$ with eigenvalues $\omega < -\omega_0$, while $W_0$ denotes the eigenspace with eigenvalues $-\omega_0 < \omega < \omega_0$). That is:
\begin{align}
    & \langle y | \tau_z |x\rangle  = 0 \quad \text{ for all $y$ in $W_0$ and $W_-$}
\label{wv2}
\end{align}
Indeed, Eq.~(\ref{wv2}) follows from the fact that the eigenvectors of $D(\vec{k})$ are orthogonal with respect to the symplectic inner product (see Sec.~\ref{Sec:background}).
Combining Eqs.~(\ref{wv1}-\ref{wv2}), we see that $x$ is orthogonal to \emph{all} the eigenspaces of $D(\vec{k})$. This is impossible since the eigenspaces of $D(\vec{k})$ span the whole $2N$ dimensional space, by our dynamic stability assumption. We conclude that there does not exist any nonzero $x \in W_+$ with $\bar{P}_+ \tau_z \bar{P}_+ |x\rangle = 0$, and therefore $W_{++}$ and $W_{+-}$ span all of $W_+$. This completes our proof that the mode separation procedure outlined in Sec.~\ref{subsec:mode} has all the required properties.

\section{Mapping in terms of Hamiltonian\label{App:Hamiltonian_Map}}
In this appendix, we derive an explicit formula for $\tilde{H}$ in terms of $H$. This formula reduces to Eq.~(\ref{Qformposdef}) in the main text in the special case where $H$ is positive definite. 

We start by writing down the (band-flattened) Hamiltonian $H$ in Eq.~(\ref{Eq:Hform}) in the BdG form:
\begin{align}
  H &=\frac{1}{2} \sum_{\vec{k}} \phi_{\vec{k}}^\dagger \, h(\vec{k}) \, \phi_{\vec{k}} \, ,
\end{align}
Here $\phi_{\vec{k}} = (b_{1,\vec{k}},\, ...,\,b_{N,\vec{k}},b^{\dagger}_{1,-\vec{k}}\, ...,\,b^{\dagger}_{N,-\vec{k}}  )^T$ and $h(\vec{k})$ is the BdG matrix. Comparing this expression with the diagonalized form
\begin{align}
        H(\vec{k}) &= \frac{1}{2} \sum_{\vec{k}} \sum_{\sigma = \pm} \sum^{R_\sigma}_{m=1}\sigma [A^{\dagger}_{m,\vec{k},\sigma} A_{m,\vec{k},\sigma}
    + A_{m,-\vec{k},\sigma} A^{\dagger}_{m,-\vec{k},\sigma}]
\end{align}
and using Eq.~(\ref{Eq:mode_annihilation_flat}), we deduce that $h(\vec{k}) = h_+(\vec{k}) - h_{-}(\vec{k})$,
with
\begin{widetext}
\begin{align}     
     & h_{\sigma}(\vec{k}) = \begin{pmatrix} u^\dagger_{\sigma}(\vec{k}) u_{\sigma}(\vec{k}) +  v^T_{\sigma}(-\vec{k}) v^*_{\sigma}(-\vec{k}) & u^\dagger_{\sigma}(\vec{k}) v_{\sigma}(\vec{k}) + v^T_{\sigma}(-\vec{k}) u^*_{\sigma}(-\vec{k}) \\ v^\dagger_{\sigma}(\vec{k}) u_{\sigma}(\vec{k}) + u^T_{\sigma}(-\vec{k}) v^*_{\sigma}(-\vec{k}) & v^\dagger_{\sigma}(\vec{k}) v_{\sigma}(\vec{k}) + u^T_{\sigma}(-\vec{k}) u^*_{\sigma}(-\vec{k}) \end{pmatrix} \, .
\label{hform}
\end{align}
\end{widetext}

Next consider the number conserving Hamiltonian $\tilde{H} = \frac{1}{2} \sum_{\vec{k}} \phi_{\vec{k}}^\dagger \, \tilde{h}(\vec{k}) \, \phi_{\vec{k}}$. By the definition of the $H \rightarrow \tilde{H}$ mapping, the BdG matrix $\tilde{h}(\vec{k})$ is given by just setting $v =0$ in $h(\vec{k})$, i.e. $\tilde{h}(\vec{k}) = \tilde{h}_+(\vec{k}) - \tilde{h}_-(\vec{k})$
where
\begin{align}
   & \tilde{h}_\sigma(\vec{k}) =  \begin{pmatrix} u^\dagger_{\sigma}(\vec{k}) u_{\sigma}(\vec{k})  & 0 \\ 0 &  u^T_{\sigma}(-\vec{k})u^{\ast}_{\sigma}(-\vec{k})   \end{pmatrix}\, ,
\end{align}
Our goal is to find an explicit formula that expresses $\tilde{h}(\vec{k})$ in terms of $h(\vec{k})$. 

To this end, it is useful to consider the dynamical matrix corresponding to $H$, which we denote by $D^{\mathrm{flat}}(\vec{k})$, to emphasize the flatnes of its bands. By definition,
\begin{align}
D^{\mathrm{flat}}(\vec{k}) =  \tau_z h (\vec{k}) 
%= \bar{h}_{d+} (\vec{k}) - \bar{h}_{d-}(\vec{k})\, .   
\end{align}
Plugging in (\ref{hform}), we derive
\begin{align}
    & D^{\mathrm{flat}}(\vec{k}) = P_{++}(\vec{k}) +P_{+-}(\vec{k}) - P_{-+}(\vec{k}) - P_{--}(\vec{k})\, ,
\label{hdflatform}
\end{align}
where
%\begin{widetext}
%\begin{align}
%    & P_{\sigma}(\vec{k}) = \begin{pmatrix} u^\dagger_{\sigma}(\vec{k}) u_{\sigma}(\vec{k}) -  v^T_{-\sigma}(-\vec{k}) v^*_{-\sigma}(-\vec{k}) & u^\dagger_{\sigma}(\vec{k}) v_{\sigma}(\vec{k}) - v^T_{-\sigma}(-\vec{k}) u^*_{-\sigma}(-\vec{k}) \\ -v^\dagger_{\sigma}(\vec{k}) u_{\sigma}(\vec{k}) + u^T_{-\sigma}(-\vec{k}) v^*_{-\sigma}(-\vec{k}) & -v^\dagger_{\sigma}(\vec{k}) v_{\sigma}(\vec{k}) + u^T_{-\sigma}(-\vec{k}) u^*_{-\sigma}(-\vec{k}) \end{pmatrix}\, ,
%\end{align}
%\end{widetext}
%It is easy to check that $P_\pm = P_\pm(\vec{k})$ are orthogonal projections, i.e.
%\begin{align}
%    & P^2_{\sigma} = P_{\sigma}\, , \quad P_{+} P_{-} = P_{-} P_{+} = 0\, .
%\end{align}
%These identities are to be expected since $P_\pm$ are spectral projectors which project onto the eigenspaces of $h^{\mathrm{flat}}_d$ with eigenvalues $\omega = \pm 1$.

%The projection matrices $P_\pm$ can further be decomposed into subspace projection matrices, such that
%\begin{align}
%    & P_{\sigma} = P_{\sigma,\sigma} + P_{\sigma,-\sigma}\, ,
%\end{align}
\begin{align}
    & P_{\sigma \sigma}(\vec{k}) = \begin{pmatrix} u^\dagger_{\sigma}(\vec{k}) u_{\sigma}(\vec{k})  & u^\dagger_{\sigma}(\vec{k}) v_{\sigma}(\vec{k}) \\ -v^\dagger_{\sigma}(\vec{k}) u_{\sigma}(\vec{k}) & -v^\dagger_{\sigma}(\vec{k}) v_{\sigma}(\vec{k})  \end{pmatrix}\, , \notag\\
    &  P_{-\sigma \sigma}(\vec{k}) = \begin{pmatrix}  -  v^T_{\sigma}(-\vec{k}) v^*_{\sigma}(-\vec{k}) &  - v^T_{\sigma}(-\vec{k}) u^*_{\sigma}(-\vec{k}) \\   u^T_{\sigma}(-\vec{k}) v^*_{\sigma}(-\vec{k}) &  u^T_{\sigma}(-\vec{k}) u^*_{\sigma}(-\vec{k}) \end{pmatrix}\, .
\end{align}
It is easy to check that $P_{\sigma \sigma'}$ are orthogonal projection matrices, i.e.
\begin{align}
    P_{\sigma \tau} P_{\sigma' \tau'} = \delta_{\sigma \sigma'} \delta_{\tau \tau'} P_{\sigma \tau}
    \end{align}
The above identities follow from the orthogonality properties of $u,v$, summarized in Eq.~(\ref{Eq:mode_orth1}). 

To understand the physical interpretation of $P_{\sigma \sigma'}$, first notice that Eq.~(\ref{hdflatform}) implies that
\begin{align}
P_{++} + P_{+-} = P_+, \quad \quad P_{-+} + P_{--} = P_-
\label{ppmsumid}
\end{align}
where $P_\pm$ are spectral projectors onto the eigenspaces $W_\pm$ of $D^{\mathrm{flat}}$ with eigenvalues $\pm 1$. We can then interpret the matrix $P_{++}$ as a projection onto the subspace $W_{++} \subset W_+$ consisting of particle-like modes while $P_{+-}$ is a projection onto the subspace $W_{+-} \subset W_+$ consisting of hole-like modes. Likewise, $P_{-+}$ is a projection onto the hole-like subspace $W_{-+} \subset W_{-}$ while $P_{--}$ is a projection onto the particle-like subspace $W_{--} \subset W_-$.

%We can now decompose the dynamical matrix using the subspace decomposition:
%\begin{align}
%   & \bar{h}_{d+}(\vec{k}) = P_{+,+}(\vec{k})- P_{-,+}(\vec{k})\, , \notag\\
%   & \bar{h}_{d-}(\vec{k}) = P_{-,-}(\vec{k})- P_{+,-}(\vec{k})\, , 
%\end{align}

To proceed further, we compare the matrix representation of $\tilde{h} = \tilde{h}(\vec{k})$ with $P_{\sigma \sigma'} = P_{\sigma \sigma'}(\vec{k})$ to derive
\begin{align}
   \tilde{h} &= \frac{\mathbbm{1}+\tau_z}{2} (P_{++} - P_{--}) \frac{\mathbbm{1}+\tau_z}{2}  \nonumber \\
   &+ \frac{\mathbbm{1}-\tau_z}{2} (P_{-+} - P_{+-}) \frac{\mathbbm{1}-\tau_z}{2} \nonumber
\label{tildeh0}
\end{align}
Next, we use (\ref{ppmsumid}) to rewrite $\tilde{h}$ as
\begin{align}
\tilde{h}   &= \frac{\mathbbm{1}+\tau_z}{2} (P_{+} -P_{+-} - P_{--}) \frac{\mathbbm{1}+\tau_z}{2}  \nonumber \\
   &+ \frac{\mathbbm{1}-\tau_z}{2} (P_{-} - P_{+-} - P_{--}) \frac{\mathbbm{1}-\tau_z}{2}
   \end{align}
To simplify this expression further, we use the fact that $P_+$ and $P_-$ are spectral projectors to derive
\begin{align}
    P_{\pm} &= \frac{1}{2}([D^{\mathrm{flat}}]^2 \pm D^{\mathrm{flat}}) \nonumber \\
    &= \frac{1}{2}(\tau_z h \tau_z h \pm \tau_z h)
    \label{Ppmhform}
\end{align}
Substituting (\ref{Ppmhform}) into (\ref{tildeh0}), we obtain
   \begin{align}
       \tilde{h} &= \frac{\mathbbm{1}+\tau_z}{2} \left(\frac{1}{2}\tau_z h \tau_z h + \frac{1}{2}\tau_z h -P_{+-} - P_{--}\right) \frac{\mathbbm{1}+\tau_z}{2}  \nonumber \\
   &+ \frac{\mathbbm{1}-\tau_z}{2} \left(\frac{1}{2} \tau_z h \tau_z h - \frac{1}{2} \tau_z h - P_{+-} - P_{--}\right) \frac{\mathbbm{1}-\tau_z}{2} \label{tildehform1}
\end{align}

The last step is to write down an explicit formula for $P_{+-}$ and $P_{--}$ in terms of $h$. To this end, we use the definition of $W_{+-}$ from Eq.~(\ref{Wpmdef}), to derive
\begin{align}
    P_{+-} &= \Theta(-\bar{P}_+ \tau_z \bar{P}_+)
    \label{Ppmtheta}
\end{align}
where
\begin{align}
    \bar{P}_+ = \Theta(P_+ P_{+}^\dagger)
\label{ppbarform}
\end{align}
Here we have introduced a new piece of notation: for any Hermitian matrix $A$, we define $\Theta(A)$ to be the matrix obtained by applying the Heaviside step function to each of the eigenvalues of $A$. That is, $\Theta(A) = \sum_a \Theta(a) |a\rangle \langle a|$ where $A = \sum_a a |a\rangle \langle a|$ is the diagonal form of $A$ and where 
\begin{equation}
    \Theta(x) = \begin{cases} 0 & x \leq 0 \\
    1 & x > 0
    \end{cases}
\end{equation}
In the same way, we have
    \begin{align}
    P_{--} &= \Theta(\bar{P}_- \tau_z \bar{P}_-)
\label{Pmmtheta}
\end{align}
where
\begin{align}
    \bar{P}_- = \Theta(P_- P_{-}^\dagger)
\label{pmbarform}
\end{align}

Substituting (\ref{Ppmtheta}) and (\ref{Pmmtheta}) into (\ref{tildehform1}) and simplifying, we obtain:
\begin{align}
\tilde{h} &= \frac{1}{4}(h + \tau_z h \tau_z + \tau_z h \tau_z h + h \tau_z h \tau_z) \nonumber \\
   & - \frac{1}{2} [\Theta(-\bar{P}_{+} \tau_z \bar{P}_+) + \Theta(\bar{P}_{-} \tau_z \bar{P}_-)] \nonumber \\
&-\frac{1}{2} \tau_z[\Theta(-\bar{P}_{+} \tau_z \bar{P}_+) + \Theta(\bar{P}_{-} \tau_z \bar{P}_-)] \tau_z 
\label{tildehform2}
\end{align}   
We are now finished: Eq.~(\ref{tildehform2}) is our desired explicit formula for $\tilde{h}$ in terms of $h$. Here, the last four terms can be expressed in terms of $h$ using Eqs.~(\ref{ppbarform}) and (\ref{pmbarform}) together with Eq.~(\ref{Ppmhform}). 

To complete our analysis, we now discuss the special case where $H$ is positive definite. In this case, $W_{+-} = 0$ since there are no hole-like modes with positive frequency. Likewise $W_{--} = 0$ since there are no particle-like modes with negative frequency. It follows that the last four terms in (\ref{tildehform2}) are $0$, since they originate from the projectors $P_{+-}$ and $P_{--}$. Thus Eq.~(\ref{tildehform2}) reduces to Eq.~(\ref{Qformposdef}) in the main text, as we wished to show.

%   P_{-,+} + P_{+,-}) \notag\\
%   &\hspace{1cm} + \tau_z (P_{+,+} - P_{-,-} - P_{-,+} + P_{+,-}) \tau_z \notag\\
%   & \hspace{1cm} + \tau_z (P_{+,+} - P_{-,-} + P_{-,+} - P_{+,-}) \notag\\
%   &\hspace{1cm} + (P_{+,+} - P_{-,-} + P_{-,+} - P_{+,-}) \tau_z]\, \notag\\
%   & = \frac{1}{4} [\tau_z \bar{h} + \bar{h}\tau_z + (\bar{h}_{+} + \bar{h}_{-})\tau_z\bar{h} + \tau_z(\bar{h}_{+} + \bar{h}_{-})\tau_z\bar{h}\tau_z]\, .
%\end{align}

%\gaurav{Further simplification needed}
%\section{Relation to second order perturbation theory}\label{app:Pairing_LL}

\section{Breakdown of mapping for fermionic systems\label{app:fermions}}
In this appendix, we discuss the fermionic analog of the $H \rightarrow \tilde{H}$ map. We also discuss the analog of the interpolating Hamiltonian $H_\epsilon$, which connects $H$ and $\tilde{H}$. We show that both the map and the interpolation are problematic because the Hamiltonian  $\tilde{H}$ does not necessarily have a spectral gap in the fermionic case.

The starting point for our discussion is a general flat band quadratic fermion Hamiltonian. Specifically, we consider a fermion Hamiltonian with $2M$ bands with frequency $\omega = 0$ and $2N-2M$ bands with frequency $\omega = \pm 1$. Similarly to Eq.~(\ref{Eq:Hform}), we can write any Hamiltonian of this kind in the diagonalized form
%unlike the bosonic case, the Hamiltonian $\tilde{H}$ can be gapless   are both problematic because the when applied to quadratic fermionic system. The fermionic BdG matrix is particle-hole symmetric and hence never positive definite.Moreover, all the fermionic normal mode simply follow the orthogonal properties of the usual inner product.Hence, there is no positive-negative norm separation.The general flat band quadratic Fermionic Hamiltonian is thus
\begin{align}
     & H = \sum_{\vec{k}} \sum^{M}_{m = 1} F^{\dagger}_{m,\vec{k}} F_{m,\vec{k}}\, ,
\end{align}
where $F_{m,\vec{k}}$ are canonical fermionic annihilation operators of the form
\begin{align}
     & F_{m,\vec{k}} = \sum^{N}_{n=1} (u_{mn} (\vec{k}) f_{n,\vec{k}} + v_{mn}(\vec{k}) f^{\dagger}_{n,-\vec{k}})\, ,
\end{align}
and where $f_{n,\vec{k}}$ is the fermion annihilation operator on sublattice $n$ and momentum $\vec{k}$. Note that, unlike the bosonic case in Eq.~(\ref{Eq:Hform}), we do not need to include terms with a negative coefficient, e.g. $-F^\dagger_{m,\vec{k}} F_{m, \vec{k}}$. The reason that we can omit these terms is that we can always make all coefficients positive by renaming $F \leftrightarrow F^\dagger$ as necessary.

Note that the coefficients $u_{mn}, v_{mn}$ obey the following identities which are consequences of the fermionic commutation relations between $F_{m,\vec{k}}, F^{\dagger}_{m,\vec{k}}$:
\begin{subequations}
\begin{align}
     & u(\vec{k}) u^{\dagger}(\vec{k}) + v(\vec{k}) v^{\dagger}(\vec{k}) = \mathbbm{1}_{M}\, ,\\
     & u(\vec{k}) v^T(-\vec{k}) + v(\vec{k}) u^T(-\vec{k}) = 0\, .
\end{align}
\end{subequations} 

With this setup, we can now state the fermionic analog of the $H \rightarrow \tilde{H}$ map:
\begin{align}
    &\tilde{H} = \sum_{\vec{k}} \sum_{m=1}^N \sum_{N=1}^N Q_{mn}(\vec{k}) f_{m,\vec{k}}^\dagger f_{n,\vec{k}} \nonumber \\
    &Q(\vec{k}) = u^\dagger(\vec{k}) u(\vec{k})
\end{align}
Likewise, the fermionic analog of the interpolating Hamiltonian $H_\epsilon$ is
\begin{align}
    & H_{\epsilon} = \sum_{\vec{k}} \sum^{M}_{m = 1} F^{\dagger}_{m,\vec{k}}(\epsilon) F_{m,\vec{k}}(\epsilon)\, ,
\end{align} 
where
\begin{align}
    & F_{m,\vec{k}}(\epsilon) = \sum^{N}_{n=1} (u_{mn} (\vec{k},\epsilon) f_{n,\vec{k}} + v_{mn}(\vec{k},\epsilon) f^{\dagger}_{n,-\vec{k}})\, ,
\end{align}
and 
\begin{align}
    & u(\vec{k},\epsilon) = u(\vec{k})\, , \quad v(\vec{k},\epsilon) = (1-\epsilon) v(\vec{k})\, ,
\end{align}

We now proceed to analyze the spectrum of the interpolating Hamiltonian $H_\epsilon$. This will also tell us the spectrum of $\tilde{H}$ since $\tilde{H}$ corresponds to the special case, $\epsilon = 1$.

To find the spectrum of $H_\epsilon$, we note that 
\begin{subequations}
\begin{align}
    & \{ F_{m,\vec{k}}(\epsilon),\, F_{m',\vec{k}'} (\epsilon) \} = 0\, ,\\
    & \{ F_{m,\vec{k}}(\epsilon),\, F^{\dagger}_{m',\vec{k}'} (\epsilon) \} = \beta_{mm'}(\vec{k},\epsilon) \delta_{\vec{k},\vec{k}'}\, , 
\end{align}
\end{subequations}
where the coefficient $\beta_{mm'}(\vec{k},\epsilon)$ is an element of the $M\times M$ matrix 
\begin{align}
    & \beta(\vec{k}, \epsilon) = u(\vec{k},\epsilon) u^{\dagger}(\vec{k},\epsilon) + v(\vec{k},\epsilon) v^{\dagger}(\vec{k},\epsilon)\, .
\end{align}
It follows from the above commutation relations that
\begin{align}
    [F_{m, \vec{k}}(\epsilon), H_\epsilon] = \sum_{m'=1}^M \beta_{mm'}(\vec{k},\epsilon) F_{m', \vec{k}}(\epsilon)
\label{FHeis}
\end{align}
Using Eq.~(\ref{FHeis}) we can read off the Heisenberg equations of motion for $F_{m,\vec{k}}(\epsilon)$ and thereby derive the normal mode spectrum of $H_\epsilon$. Following this line of reasoning, one can show that the nonzero eigenvalues of $H_\epsilon$ are identical to the eigenvalues of the matrix $\beta(\vec{k},\epsilon)$, up to a $\pm$ sign. 

All that remains is to determine the eigenvalue spectrum of $\beta(\vec{k},\epsilon)$. To this end, we simplify $\beta$ as
\begin{align}
     & \beta(\vec{k},\epsilon) = \mathbbm{1}_{M} - \epsilon (2-\epsilon) v(\vec{k})v^{\dagger}(\vec{k})\, .
\end{align}
Since $v(\vec{k})v^{\dagger}(\vec{k})$ is non-negative definite, we can see that the energy gap for $\epsilon > 0$ is in general less than $1$, i.e. the gap is \emph{smaller} than the original flat band Hamiltonian. (This should be contrasted with the bosonic interpolation, where the every gap is always \emph{larger} than the original flat band system). At the same time, we can also see that the energy gap remains open for any $\epsilon < 1$ since the eigenvalues of $v(\vec{k})v^{\dagger}(\vec{k})$ are $\leq 1$. In other words, the fermionic interpolation is valid for $\epsilon < 1$. 

The place where the interpolation can break down is at the endpoint $\epsilon = 1$ (corresponding to $\tilde{H}$): at this endpoint, the energy gap can close, as we will demonstrate in an example below. This (possible) closing of the spectral gap at $\epsilon = 1$ is the key point of this appendix, and it is the fundamental reason that our mapping cannot be applied to general fermionic systems.

To show such a gap closing is indeed possible for a fermionic system, we consider the (flat band) Kitaev chain Hamiltonian under periodic boundary conditions. The BdG matrix for this system is:
\begin{align}
     & h_{BdG}(k) = h_y(k)\sigma_y + h_z(k)\sigma_z \, , 
\end{align}
where $h_y(k) = \sin k$ and $h_z(k) = -\cos k$. 
The $u$ and $v$ coefficients for this Hamiltonian are
\begin{align}
    u(k) = \sin(k/2), \quad \quad v(k) = i \cos(k/2)
\end{align}
%h_z/\sqrt{h^{2}_y+h^2_z}$, the positive and negative energy eigenvectors are respectively
%\begin{subequations}
%\begin{align}
%     & u_{+}(k) = \biggl ( \textup{e}^{i\phi_{k}}\cos\frac{\theta_{k}}{2},\, \sin\frac{\theta_k}{2} \biggr )^T\, ,\\
%     & u_{-}(k) = \biggl (-\textup{e}^{i\phi_{k}}\sin\frac{\theta_{k}}{2},\, \cos\frac{\theta_k}{2} \biggr )^T\, .
%\end{align}
%\end{subequations}
If we carry out the above mentioned interpolation under the parameter $\epsilon\in [0,1]$, we find
\begin{align}
    & \beta(k, \epsilon) = \sin^2(k/2) + (1-\epsilon)^2 \cos^2(k/2).
\end{align}
We can see that when $\epsilon = 1$, we have $\beta = 0$ at $k = 0$, signifying that the gap closes at the endpoint of the interpolation. This gap closing is expected, of course, since it is impossible to interpolate between the Kitaev chain and a number conserving system without closing the gap.

%\section{Symmetries}
%\gaurav{Let's discuss if we want to add this today and its structure.}
%%%%%%%%%%%%%%%%%%%%%%%%%%%%%%%%%%%%%%%%%%%%%%%%%%%%%%%%%%%%%%%%%%%%%%%%%%%%%%%%%%%%%%%%%%%%%%%%%%%%

%%%%%%%%%%%%%%%%%%%%%%%%%%%%%%%%%%%%%%%%%%%%%%%%%%%%%%%%%%%%%%%%%%%%%%%%%%%%%%%%%%%%%%%%%%%%%%%%%%%%
\bibliography{bibliography}
%%%%%%%%%%%%%%%%%%%%%%%%%%%%%%%%%%%%%%%%%%%%%%%%%%%%%%%%%%%%%%%%%%%%%%%%%%%%%%%%%%%%%%%%%%%%%%%%%%%%

\end{document}